\documentclass[aps,pre,subeqns.floatfix,amsmath,twocolumn]{revtex4}
\usepackage[pdftex]{graphicx}
\usepackage{amsmath}
\usepackage{amsfonts}
\usepackage{graphicx}
\usepackage{hyperref}
\usepackage{comment}
\usepackage{mathtools}
\usepackage{verbatim}
\usepackage{enumerate}
\usepackage[outdir=./]{epstopdf}
\usepackage{subfigure}
\usepackage{hyperref}
\usepackage{latexsym}
\usepackage{dcolumn}
\usepackage{epsf}
\usepackage{float}

\usepackage{subfigure}
\DeclarePairedDelimiter\abs{\lvert}{\rvert}

\usepackage{color}
\usepackage[utf8]{inputenc}

\begin{document}
	\title{Double explosive transitions in adaptive multilayer networks with higher-order interactions}
	\author{Anath Bandhu Das}
	\email{anathbandhu0498@gmail.com}
	\author{Pinaki Pal}
	\email{ppal.maths@nitdgp.ac.in}
	\affiliation{Department of Mathematics, National Institute of Technology, Durgapur~713209, India}

\begin{abstract}
Can asymmetry between two interacting networks fundamentally change how they synchronize? We identify double explosive transitions in the forward direction, backward direction, or a combination thereof, with single or double hysteresis loops in an adaptive bilayer multiplex network of Kuramoto oscillators with pairwise and three-body interactions and asymmetric phase lags. Using the Ott-Antonsen reduction, we derive a low-dimensional system and perform a stability analysis. The reduced model accurately captures the full microscopic dynamics and enables analytical expressions for the bifurcation points. Systematic mapping across multiple parameter planes reveals eight distinct synchronization regimes. The relative ordering of saddle-node and pitchfork bifurcation points---controlled by phase-lag asymmetry, cross-layer adaptation, and the higher-order interaction strength---creates two distinct coherent branches (weak and strong), giving rise to double explosive transitions. Crucially, phase-lag asymmetry acts as a robust control knob: while symmetric phase lags suppress explosive transitions, layer-specific differences promote multistability and double explosive transitions. The higher-order interaction strength $K_2$ and adaptation strengths $q$, $p$, and $h$ further modulate these transitions in a complex, parameter-dependent manner. Excellent agreement between analytical predictions and numerical simulations confirms the robustness of our reduced description.
\end{abstract}

\maketitle


\section{Introduction}
Over the past two decades, network science has become an important framework for studying the emergence of collective phenomena in complex systems. By modeling the interactions among the individual components of such systems as networks or graphs~\cite{strogatz2001exploring,newman2003structure}, researchers can examine how the underlying connectivity influences the resulting dynamical behavior~\cite{boccaletti2006complex}. A broad spectrum of processes has been investigated within this framework, including percolation~\cite{christensen2005complexity}, synchronization~\cite{pikovsky2003universal,arenas2008synchronization}, epidemic spreading~\cite{pastor2015epidemic}, and cooperative behavior~\cite{szabo2007evolutionary}. This extensive research on the relationship between network structure and dynamical function has revealed a range of microscopic mechanisms through which collective states can arise, evolve, and be regulated.

One of the most significant phenomena studied in this context is the emergence of explosive transitions, characterized by abrupt changes in macroscopic states. Such transitions can arise from microscopic dynamical mechanisms shaped by the underlying network topology, which may hinder the gradual development of collective behavior~\cite{boccaletti2016explosive,d2019explosive}. Explosive transitions have been widely observed in synchronization dynamics~\cite{gomez2011explosive,leyva2012explosive,arola2022emergence}, where collective coherence can change abruptly with the coupling strength. Such behavior can result from structural heterogeneity in the network combined with a positive correlation between the natural frequencies of oscillators and their degrees~\cite{gomez2011explosive,kundu2017transition}. These conditions have subsequently been extended to partial frequency-degree correlations~\cite{pinto2015explosive,kundu2019synchronization} and frequency-weighted coupling~\cite{zhang2013explosive,leyva2013explosive,xu2016synchronization}. Other mechanisms capable of inducing explosive transitions include local adaptive strategy in mono- and multilayer networks~\cite{zhang2015explosive,danziger2016explosive,khanra2018explosive,khanra2021explosive}, time-delayed interactions~\cite{peron2012explosive,kachhvah2019delay}, and higher-order network structures~\cite{skardal2020higher,laptyeva2025explosive,malizia2025hyperedge}.

Beyond these structural and dynamical mechanisms, the phase-lag parameter in the Sakaguchi-Kuramoto model~\cite{sakaguchi1986soluble} provides an intrinsic means of controlling synchronization transitions by breaking the odd symmetry of the coupling function. The transition to synchronization can become highly nontrivial when phase lag interacts with the coupling function, frequency distribution, and network topology~\cite{omel2012nonuniversal,omel2013bifurcations,omel2016there}. In certain regimes, the order parameter may develop a weakly coherent state before undergoing an abrupt jump in the order parameter, leading to tiered synchronization~\cite{skardal2022tiered,rajwani2023tiered,manoranjani2023phase,das2026effect}. Beyond tiered synchronization, multistep and stairlike transitions have been reported in a variety of networked oscillator systems. In random networks, partial correlations between node degree and natural frequency can give rise to stairlike transitions in the Sakaguchi--Kuramoto model~\cite{kundu2019synchronization}. Similar behavior has been observed in finite networks of second-order Kuramoto oscillators, where clusters of different sizes may coexist within hysteretic regimes when the inertial effect is sufficiently strong~\cite{olmi2014hysteretic}. More complex stairlike transitions have subsequently been reported in systems with strong inertia~\cite{gao2021synchronized} and in oscillator networks with higher-order interactions~\cite{carballosa2023cluster}. Furthermore, appropriate frequency selection together with temporal adjustment of network connections can induce stairlike discontinuous transitions in finite Kuramoto networks~\cite{fialkowski2023heterogeneous}. In duplex networks, a double hysteresis loop has also been quantitatively identified in the high-phase-lag regime~\cite{seif2025double}.

Higher-order interactions offer additional mechanisms for generating multistep synchronization. A combination of attractive higher-order interactions and repulsive pairwise coupling has been shown to produce a stepwise explosive transition~\cite{li2025higher}. More recently, adaptive higher-order coupling in hypergraphs has been found to produce a double explosive transition, in which partial coherence emerges gradually before the system undergoes an abrupt transition to a strongly synchronized state~\cite{dutta2025double}. In addition, suitable choices of higher-order coupling strengths, including interactions up to fifth order, have been shown to generate double jumps in synchronization transitions~\cite{costa2025exact}.

While these studies highlight several structural and dynamical mechanisms underlying multistep transitions, our previous work~\cite{das2026effect} demonstrated that phase lag can also play an important role in shaping synchronization transitions in adaptive multilayer networks with higher-order interactions. We found that phase frustration can either suppress or promote discontinuous synchronization depending on the coupling mechanism: it inhibits tiered and explosive synchronization when the triadic coupling strength is fixed, whereas it promotes discontinuous transitions when the pairwise coupling strength is fixed. However, our previous study considered symmetric phase-lag configurations, with the same phase-lag parameter assigned to all layers. This naturally raises the question of how synchronization transitions are affected when the phase-lag parameters differ between layers. The role of asymmetric phase lags, where different layers possess distinct phase-lag parameters, remains largely unexplored. Such asymmetry is physically relevant to multilayer systems in which different subsystems may exhibit distinct phase-response characteristics. For instance, synaptic delays and phase-response properties can vary across brain regions~\cite{buldu2018frequency,buzsaki2006rhythms}, while transmission delays in power-grid networks may differ across geographical regions~\cite{bottcher2020time}. Whether such layer-dependent phase lags can fundamentally modify synchronization transitions and generate transition scenarios that are inaccessible under symmetric configurations remains an open question. Understanding and controlling complex synchronization transitions is important in several real-world systems. In neural systems, abrupt transitions toward excessive synchronization can be associated with pathological activity, such as epileptic seizures~\cite{wang2017explosive,wang2017small}. Similarly, sudden loss of coherent operation in power-grid networks can contribute to large-scale failures~\cite{dobson2007complex}. Identifying dynamical mechanisms that can suppress such abrupt transitions or control their occurrence is therefore of both fundamental and practical importance.

In this work, we extend our previous framework by investigating synchronization transitions in an adaptive bilayer multiplex network of Kuramoto oscillators with pairwise and three-body interactions and asymmetric phase-lag parameters, $\beta_1\neq\beta_2$. The coupling strength within each layer is adaptively modulated by the global order parameter of the opposite layer, establishing mutual cross-layer dependence. Using the Ott-Antonsen reduction, we derive a low-dimensional description of the macroscopic dynamics and perform a comprehensive stability analysis. Our results demonstrate that the interplay of phase-lag asymmetry, cross-layer adaptation, and higher-order interactions produces a rich variety of synchronization scenarios. We show that the relative ordering of saddle-node and pitchfork bifurcation points determines whether the system exhibits single or double jumps and whether these transitions are accompanied by a single or multiple hysteresis loops. Through systematic exploration of the parameter space, we demonstrate that phase-lag asymmetry provides an effective control mechanism for switching between qualitatively different transition regimes, including multistability, tiered transitions, and double explosive transitions. The higher-order interaction strength and adaptation exponents further modulate these regimes, highlighting the intricate interplay between dynamical asymmetry, adaptive coupling, and higher-order interactions in shaping collective synchronization.


\section{Model}
We consider a bilayer multiplex network in which each layer consists of $N$ globally coupled Kuramoto oscillators. The dynamical state of each oscillator is governed by a combination of pairwise (1-simplex) and three-body (2-simplex) interactions, both of which are subject to a constant phase lag. A defining feature of the present model is the absence of conventional inter-layer links connecting individual nodes across layers. Instead, the coupling within each layer is adaptively modulated by the global synchronization state of the opposite layer, thereby establishing a mutual, one-to-one interdependence between the layers. Let $\theta_{i,l}$ and $\omega_{i,l}$ denote the instantaneous phase and the natural frequency, respectively, of the $i$-th oscillator in layer $l~(l=1,2)$. The temporal evolution of the system is governed by the following set of coupled ordinary differential equations:
\begin{equation}\label{eqn1}
   \begin{aligned}
       \dot{\theta}_{i,1}&=\omega_{i,1}+\frac{K_1\left(1+qr_{1,2}^{p_1}\right)}{N}\sum_{j=1}^N \sin(\theta_{j,1}-\theta_{i,1}-\beta_{1}) \\
     &+\frac{K_{2}\left(1+qr_{1,2}^{h_1}\right)}{N^2} \sum_{j=1}^N \sum_{k=1}^N \sin(2\theta_{j,1}-\theta_{k,1}-\theta_{i,1}-\beta_{1}),\\
        \dot{\theta}_{i,2}&=\omega_{i,2}+\frac{K_{1}\left(1+qr_{1,1}^{p_2}\right)}{N}\sum_{j=1}^N \sin(\theta_{j,2}-\theta_{i,2}-\beta_{2}) \\
     &+\frac{K_{2}\left(1+qr_{1,1}^{h_2}\right)}{N^2} \sum_{j=1}^N \sum_{k=1}^N \sin(2\theta_{j,2}-\theta_{k,2}-\theta_{i,2}-\beta_{2}),  
   \end{aligned}
\end{equation}
for $i=1,2,\dots,N$. The parameters $K_1$ and $K_2$ quantify the coupling strengths of the pairwise and three-body interactions, respectively, while $\beta_l$ denotes the corresponding phase-lag parameter in layer $l$. The global dynamical state of each layer is characterized by the Kuramoto complex order parameter
  \begin{equation}
      z_{1,l}=r_{1,l}e^{\iota \phi_{1,l}}=\frac{1}{N} \sum_{j=1}^{N}e^{\iota \theta_{j,l}}, \label{eqn2}
  \end{equation}
  where $r_{1,l}\in[0,1]$ measures the degree of phase coherence and $\phi_{1,l}$ is the corresponding mean phase angle.

  The terms involving $r_{1,2}^{p_1}, r_{1,2}^{h_1}, r_{1,1}^{p_2}, r_{1,1}^{h_2}$ encode the cross-layer feedback: the coupling amplitudes in layer 1 are modulated by the coherence $r_{1,2}$ of layer 2, and similarly, the coupling amplitudes in layer 2 are modulated by the coherence $r_{1,1}$ of layer 1. The exponents $p_l$ and $h_{l}$ control the nonlinearity of this modulation for the pairwise and three-body interactions, respectively, following a power-law adaptation scheme commonly employed in studies of adaptive dynamical systems~\cite{filatrella2007generalized,zou2020dynamics,xu2021collective,rajwani2023tiered}. The parameter $q\ge0$ governs the overall strength of the adaptation. In the limit $q=0$, the system reduces to two independent layers, each governed by the classical Kuramoto model with higher-order interactions. For $q>0$, the coupling strengths become functions of the collective state of the opposing layer, enabling controlled modulation of oscillator dynamics in response to emergent collective behavior.
  
  The natural frequencies $\omega_{i,l}$ are assumed to be quenched random variables drawn from a continuous probability distribution $g_{l}(\omega)$. We consider a unimodal frequency distribution for $g_{l}(\omega)$. Specifically, we adopt the Lorentzian distribution
     \begin{equation}
       g_{l}(\omega)=\frac{\gamma_{l}}{\pi[\gamma_{l}^2+(\omega_{l}-\omega_{0,l})^2]},~\gamma_{l}>0, \label{eqn3}
   \end{equation}
   where $\omega_{0,l}$ denotes the center frequency and $\gamma_{l}$ characterizes the half-width of the distribution for layer $l$. This distribution is symmetric about its mean $\omega_{0,l}$ and, owing to its simple pole structure in the complex plane, permits an exact reduction of the governing equations to a low-dimensional system via the Ott-Antonsen ansatz~\cite{ott2008low}. This form of frequency distribution is representative of coupled oscillator systems characterized by a single dominant frequency component, a scenario frequently encountered in many physical systems.


\section{Dimensionality reduction}
Having established the full microscopic model in the preceding section, we now derive a low-dimensional description of the collective macroscopic dynamics. Our goal is to reduce the high-dimensional system (\ref{eqn1}) to a tractable set of equations governing the evolution of the global order parameters. In addition to the first-order parameter $z_{1,l}$ defined in Eq.~(\ref{eqn2}), the presence of three-body interactions necessitates the inclusion of the second harmonic
  \begin{equation}
      z_{2,l}=r_{2,l}e^{\iota \phi_{2,l}}=\frac{1}{N} \sum_{j=1}^{N}e^{2\iota \theta_{j,l}}. \label{eqn4}
  \end{equation}
While $z_{1,l}$ characterizes the centroid of the $N$ points $e^{\iota \theta_{j,l}}$ on the unit circle and suffices for pairwise interactions, the nonlinearities introduced by the three-body couplings require the second harmonic $z_{2,l}$, which represents the centroid of the points $e^{2\iota\theta_{j,l}}$.

Using the definitions of $z_{1,l}$ and $z_{2,l}$, the governing equations (\ref{eqn1}) can be rewritten in the compact form
\begin{equation}\label{eqn5}
    \begin{aligned}
        \dot{\theta}_{i,1}&=\omega_{i,1}+\frac{1}{2\iota} \left[H_{1} e^{-\iota(\theta_{i,1}+\beta_{1})} - H_{1}^{*} e^{\iota(\theta_{i,1}+\beta_{1})} \right], \\
        \dot{\theta}_{i,2}&=\omega_{i,2}+\frac{1}{2\iota} \left[H_{2} e^{-\iota(\theta_{i,2}+\beta_{2})} - H_{2}^{*} e^{\iota(\theta_{i,2}+\beta_{2})} \right],
    \end{aligned}
\end{equation}
where
\begin{equation}\label{eqn6}
    \begin{aligned}
        H_{1}&=K_{1}z_{1,1}\left(1+qr_{1,2}^{p_1}\right)+K_{2}z_{2,1} z_{1,1}^{*}\left(1+qr_{1,2}^{h_1}\right), \\
        H_{2}&=K_{1}z_{1,2}\left(1+qr_{1,1}^{p_2}\right)+K_{2}z_{2,2} z_{1,2}^{*}\left(1+qr_{1,1}^{h_2}\right),
    \end{aligned}
\end{equation}
with $z_{1,l}^{*}$ denoting the complex conjugate of $z_{1,l}$.

In the thermodynamic limit $N\to\infty$, the state of each layer is described by a continuous probability density function $f_l(\theta,\omega,t)$, defined such that $f_l(\theta,\omega,t)d\theta d\omega$ gives the fraction of oscillators in layer $l$ with phases in $[\theta,\theta+d\theta]$ and natural frequencies in $[\omega,\omega+d\omega]$ at time $t$. Conservation of the number of oscillators requires that $f_l$ satisfies the continuity equation
 \begin{equation}
    \frac{\partial f_{l}}{\partial t}+\frac{\partial}{\partial \theta} (f_{l}v_{l})=0, \label{eqn7}
 \end{equation}
 where $v_l=\dot{\theta}_{i,l}$ is the angular velocity of an oscillator with phase $\theta$ and natural frequency $\omega$ in layer $l$. Since $f_{l}(\theta,\omega,t)$ is $2\pi$-periodic in $\theta$, it admits a Fourier series expansion of the form
    \begin{equation}
       f_{l}=\frac{g_{l}(\omega)}{2\pi} \left[1+\sum_{n=1}^\infty a_{n,l}(\omega,t)e^{\iota n\theta}+\sum_{n=1}^\infty a_{n,l}^{*}(\omega,t)e^{-\iota n\theta} \right], \label{eqn8}
   \end{equation}
   where $a_{n,l}(\omega,t)$ denotes the $n$-th Fourier coefficient for layer $l$, and $a_{n,l}^{*}(\omega,t)$ is its complex conjugate.

Now, following the Ott-Antonsen ansatz~\cite{ott2008low}, the Fourier coefficients are assumed to decay geometrically as $a_{n,l}=\alpha_{l}^{n},~\abs{\alpha_{l}}\ll1$, where $\alpha_{l}=\alpha_{l}(\omega,t)$ is a complex function of frequency and time. Substituting the Fourier expansion (\ref{eqn8}) into the continuity equation (\ref{eqn7}) and requiring consistency of the resulting hierarchy of modes yields a closed, low-dimensional evolution equation for $\alpha_l$
  \begin{equation}
       \dot{\alpha_{l}}=-\iota \omega_{l}\alpha_{l}-\frac{1}{2}\left(\alpha_{l}^2e^{-\iota\beta_{l}}H_{l}-e^{\iota\beta_{l}}H_{l}^* \right), \label{eqn9}
   \end{equation}
   with the order parameters expressed in the continuum limit as
   \begin{equation}
       z_{m,l}=\int_{-\infty}^{\infty}{\alpha_{l}^*}^m(\omega,t) g_{l}(\omega) d\omega,~m=1,2. \label{eqn10}
   \end{equation}
   Having obtained the reduced dynamics for the OA variable $\alpha_l$ in Eq.~(\ref{eqn9}), we now evaluate the frequency integrals in Eq.~(\ref{eqn10}) to obtain a closed system for the macroscopic order parameters $z_{1,l}$ and $z_{2,l}$. The integrals can be evaluated using Cauchy's residue theorem by closing the contour in the lower half of the complex $\omega$-plane. The Lorentzian distribution $g_{l}(\omega)$ in Eq.~(\ref{eqn3}) possesses a simple pole at $\omega=\omega_{0,l}-\iota\gamma_{l}$. Closing the contour via a semicircle of infinite radius in the lower half-plane, the integrals receive contributions solely from this pole. Applying the residue theorem to Eq.~(\ref{eqn10}) for $m=1$ yields $z_{1,l}=\alpha_{l}^*(\omega_{0,l}-\iota\gamma_{l},t)$. Similarly, for $m=2$, we obtain $z_{2,l}={\alpha_{l}^*}^2(\omega_{0,l}-\iota\gamma_{l},t)=z_{1,l}^2$.

   Substituting the polar forms $z_{1,l}=r_{1,l}e^{\iota \phi_{1,l}}$ and $z_{2,l}=z_{1,l}^2$ into the evolution equation for $z_{1,l}$ obtained by evaluating Eq.~(\ref{eqn9}) at the pole $\omega=\omega_{0,l}-\iota\gamma_{l}$ and taking the complex conjugate, and then separating the real and imaginary parts, yields the governing equations for the amplitude $r_l$ and phase $\phi_l$ of the order parameter. For notational simplicity, we henceforth drop the redundant subscript `1' in $r_{1,l}$ and $\phi_{1,l}$ and denote them as $r_l$ and $\phi_l$. The amplitude dynamics of the two coupled layers are given by
   \begin{equation}\label{eqn11}
   \scalebox{0.87}{$
   \begin{aligned}
       \dot{r}_{1} &= -\Delta_{1} r_{1} + \frac{r_{1}(1 - r^{2}_{1}) \cos\beta_{1}}{2} \left[ K_{1}(1 + qr_{2}^{p_{1}}) + K_{2}(1 + qr_{2}^{h_{1}}) r^{2}_{1} \right],\\
       \dot{r}_{2} &= -\Delta_{2} r_{2} + \frac{r_{2}(1 - r^{2}_{2}) \cos\beta_{2}}{2} \left[ K_{1}(1 + qr_{1}^{p_{2}}) + K_{2}(1 + qr_{1}^{h_{2}}) r^{2}_{2} \right].
   \end{aligned}
   $}
   \end{equation}
   Similarly, the corresponding phase equations are obtained as
   \begin{equation}\label{eqn12}
   	\scalebox{0.94}{$
       \begin{aligned}
           \dot{\phi}_{1}&=\omega_{0,1}-\frac{(1+r^{2}_{1})\sin\beta_{1}}{2}\left[ K_{1}(1 + qr_{2}^{p_{1}}) + K_{2}(1 + qr_{2}^{h_{1}}) r^{2}_{1} \right], \\
           \dot{\phi}_{2}&=\omega_{0,2}-\frac{(1+r^{2}_{2})\sin\beta_{2}}{2}\left[ K_{1}(1 + qr_{1}^{p_{2}}) + K_{2}(1 + qr_{1}^{h_{2}}) r^{2}_{2} \right].
       \end{aligned}
   $}
   \end{equation}
   The amplitude equations~(\ref{eqn11}) govern the evolution of the synchronization level in each layer, while the phase equations~(\ref{eqn12}) determine the corresponding mean phases. The mutual interdependence between the layers is preserved in the reduced description through the appearance of $r_2$ in the adaptive factors of layer $1$ and $r_1$ in those of layer $2$.

   We now analyze the stability of the fixed points of the reduced system~(\ref{eqn11}). Rewriting (\ref{eqn11}) as $\dot{\pmb{r}}=\pmb{F}(\pmb{r})$ with $\pmb{r}=(r_1,r_2)^{\mathrm{T}}$, the fixed points satisfy $\pmb{F}(\pmb{r})=\pmb{0}$, yielding four qualitatively different steady-state solutions, namely, the three trivial fixed points $\mathcal{S}_0=(0,0)$, $\mathcal{S}_1=(r_1^{+},0)$, and $\mathcal{S}_2=(0,r_2^{+})$, and the nontrivial fixed point $\mathcal{S}_3=(r_1^{*},r_2^{*})$. Here $r_k^{+}$ is obtained by setting the appropriate single-layer equation to zero
   \begin{equation}\label{eqn13}
   	\scalebox{0.9}{$
   		\begin{aligned}
   			 r_k^{+}=\sqrt{\frac{(K_2-K_1)\cos\beta_k\pm\sqrt{(K_1+K_2)^2\cos^2\beta_k-8K_2\cos\beta_k}}{2K_2\cos\beta_k}},
   		\end{aligned}
    $}
   \end{equation}
   provided the discriminant is positive and $r_k^{+}\in(0,1]$. The nontrivial fixed point $(r_1^{*}, r_2^{*})$ with both components nonzero is determined numerically by solving $F_1=0$ and $F_2=0$.

   The stability of a fixed point $\pmb{r}^{*}$ is determined by the eigenvalues of the Jacobian matrix
   \begin{equation*}
   \mathcal{J}(\pmb{r}^*) = \begin{pmatrix}
   \frac{\partial F_1}{\partial r_1} & \frac{\partial F_1}{\partial r_2} \\[6pt]
   \frac{\partial F_2}{\partial r_1} & \frac{\partial F_2}{\partial r_2}
   \end{pmatrix}_{\pmb{r} = \pmb{r}^*}.
   \end{equation*}
   
   For $\mathcal{S}_0=(0,0)$, the Jacobian is diagonal
   \begin{equation*}
   \mathcal{J}(\mathcal{S}_0) = \begin{pmatrix}
   -\Delta_1+\frac{K_1\cos\beta_1}{2} & 0 \\[6pt]
   0 & -\Delta_2+\frac{K_1\cos\beta_2}{2}
   \end{pmatrix},
   \end{equation*}
   with eigenvalues $\lambda_1^{(0)}=-\Delta_1+\frac{K_1\cos\beta_1}{2}$ and $\lambda_2^{(0)}=-\Delta_1+\frac{K_1\cos\beta_2}{2}$. Thus, $\mathcal{S}_0$ is stable when $K_1<\min\left(\frac{2\Delta_1}{\cos\beta_1},\frac{2\Delta_2}{\cos\beta_2}\right)$.
   
   For $\mathcal{S}_1=(r_1^{+},0)$, the Jacobian is upper triangular
   \begin{equation*}
   \mathcal{J}(\mathcal{S}_1) = \begin{pmatrix}
   J_{11}^{(1)} & J_{12}^{(1)} \\[6pt]
   0 & J_{22}^{(1)}
   \end{pmatrix},
   \end{equation*}
   where 
   \begin{equation*}
   	\scalebox{0.9}{$
       \begin{aligned}
           J_{11}^{(1)}=-\Delta_1&+\frac{\cos\beta_1}{2}\left[K_1\left(1-3(r_1^{+})^2\right)+K_2\left(3(r_1^{+})^2-5(r_1^{+})^4\right)\right],\\
           J_{22}^{(1)}=-\Delta_2&+\frac{K_1\cos\beta_2}{2}\left[1+q(r_1^{+})^{p_2}\right].
        \end{aligned}
    $}
   \end{equation*}
   The eigen values are given by $J_{11}^{(1)}$ and $J_{22}^{(1)}$. Thus, $\mathcal{S}_1$ is stable when $\max\left(J_{11}^{(1)},J_{22}^{(1)}\right)<0$.
   
   For $\mathcal{S}_2=(0,r_2^{+})$, the Jacobian is lower triangular
   \begin{equation*}
   \mathcal{J}(\mathcal{S}_2) = \begin{pmatrix}
   J_{11}^{(2)} & 0 \\[6pt]
   J_{21}^{(2)} & J_{22}^{(2)}
   \end{pmatrix},
   \end{equation*}
   where
   \begin{equation*}
   	\scalebox{0.9}{$
       \begin{aligned}
           J_{11}^{(2)}=-\Delta_1&+\frac{K_1\cos\beta_1}{2}\left[1+q(r_2^{+})^{p_1}\right],\\
            J_{22}^{(2)}=-\Delta_2&+\frac{\cos\beta_2}{2}\left[K_1\left(1-3(r_2^{+})^2\right)+K_2\left(3(r_2^{+})^2-5(r_2^{+})^4\right)\right].
       \end{aligned}
    $}
   \end{equation*}
   The eigen values are given by $J_{11}^{(2)}$ and $J_{22}^{(2)}$. Thus, $\mathcal{S}_2$ is stable when $\max\left(J_{11}^{(2)},J_{22}^{(2)}\right)<0$.
   
   For $\mathcal{S}_3=(r_1^{*},r_2^{*})$, the stability cannot be determined analytically, as the Jacobian eigenvalues do not admit closed-form expressions in terms of the system parameters.


\begin{figure*}
	\centering
	\includegraphics[width=\linewidth]{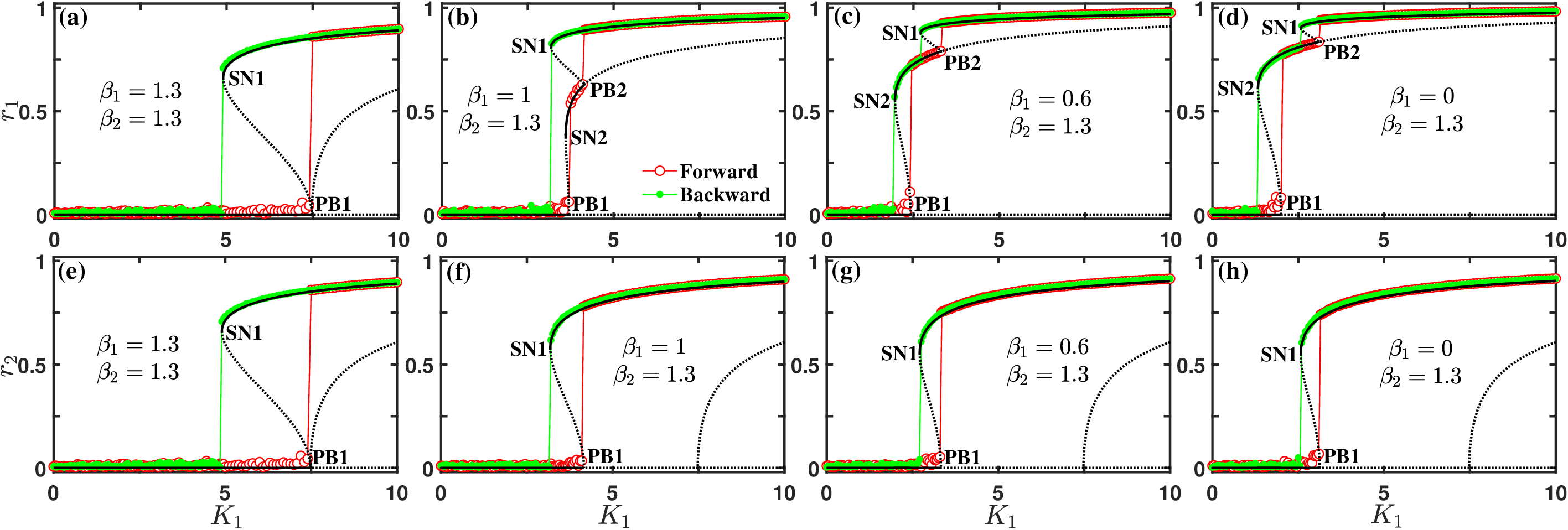}
	\caption{Evolution of the layer order parameters with $K_1$ under adiabatic forward (red, open circles) and backward (green, filled circles) continuation, together with the stable (black solid) and unstable (black dotted) branches from the reduced-order model, for $K_{2}=5$, $p_1=p_2=2$, $h_1=h_2=2$, $q=2$, and $\beta_2=1.3$. Top row (a)-(d): first-layer order parameter $r_1$; bottom row (e)-(h): second-layer order parameter $r_2$. Columns share a common $\beta_1$: $1.3$ (a),(e); $1$ (b),(f); $0.6$ (c),(g); and $0$ (d),(h). Panels (a),(e), where $\beta_1=\beta_2$, are the symmetric reference case; the remaining columns make the phase lag increasingly asymmetric by lowering $\beta_1$ at fixed $\beta_2$. In the first layer this asymmetry changes the transition scenario: (a) shows a single explosive jump with hysteresis; in (b) the forward branch acquires a second jump while the backward branch still jumps once; in (c) and (d) both branches jump twice, producing two separate hysteresis loops. The second layer, by contrast, stays singly explosive throughout (e)-(h), with the hysteresis width narrowing from (e) to (g) and remaining essentially unchanged between (g) and (h). The comparison shows that an asymmetric layer-specific phase lag can push one layer into a double explosive transition while leaving the other conventionally explosive.}
	\label{fig1}
\end{figure*}


\section{Results}
We now present the numerical and analytical results for the reduced system (\ref{eqn11}). The full microscopic equations (\ref{eqn1}) are integrated using the Euler method with a time step of $0.001$ over a total simulation time of $20^4$. The natural frequencies are drawn from a Lorentzian distribution with $\omega_{0,l}=0$ and $\gamma_l=1$ for both layers. The number of oscillators per layer is fixed at $N=10^4$. We consider symmetric adaptation exponents $p_1=p_2=p$ and $h_1=h_2=h$, while the phase-lag parameters $\beta_l$ are chosen asymmetrically $(\beta_1\neq\beta_2)$ within the interval $[0,\frac{\pi}{2}]$ unless stated otherwise. Initial conditions are sampled uniformly from $\theta\in[0,2\pi]$. Adiabatic continuation in the coupling parameter $K_1$ is performed by using the final state of the previous run as the initial condition for the next. Stable and unstable branches are identified from the eigenvalue analysis of the Jacobian of Eq.~(\ref{eqn11}). In the figures, solid black lines denote stable analytical branches, dotted black lines denote unstable branches, red curves with open circles represent forward continuation, and green curves with filled circles represent backward continuation. The numerical results show excellent agreement with the analytical predictions.

\begin{figure*}
	\centering
	\includegraphics[width=\linewidth]{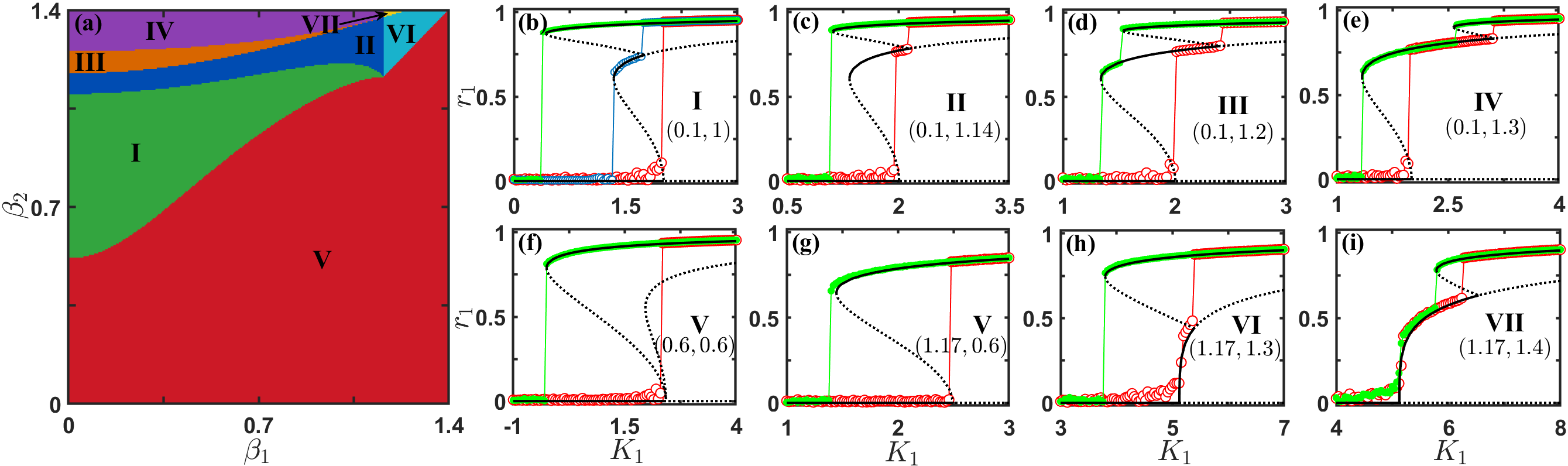}
	\caption{Classification of the $K_1$-induced transition scenarios of the first-layer order parameter $r_1$ across the phase-lag plane, for fixed $K_{2}=5$, $p_1=p_2=2$, $h_1=h_2=2$, and $q=2$. (a) Phase diagram in the $\beta_1\text{-}\beta_2$ plane, showing seven distinct dynamical regimes: I (forest green) multiple transitions with initial-condition dependence; II (navy blue) a forward double jump and a backward single jump; III (burnt orange) forward and backward double jumps with a single hysteresis structure; IV (violet) forward and backward double jumps with two distinct hysteresis loops; V (crimson red) classical explosive synchronization; VI (teal) tiered transition of type I; VII (gold) tiered transition of type II. (b)-(i) Representative $r_1\text{-}K_1$ diagrams at $(\beta_1,\beta_2)=(0.1,1)$ [I], $(0.1,1.14)$ [II], $(0.1,1.2)$ [III], $(0.1,1.3)$ [IV], $(0.6,0.6)$ [V], $(1.17,0.6)$ [V], $(1.17,1.3)$ [VI], $(1.17,1.4)$ [VII], respectively. Black solid/dotted: stable/unstable branches from the reduced-order model; red open circles/green filled circles: forward/backward adiabatic continuation; light blue open circles in (b): additional forward route from a different initial condition.}
	\label{fig2}
\end{figure*}

Figure \ref{fig1} shows the evolution of the global order parameters $r_1$ and $r_2$ with $K_1$ for fixed $K_2=5$, $p_1=p_2=2$, $h_1=h_2=2$, $q=2$, $\beta_2=1.3$, and four values of $\beta_1$: $1.3$, $1$, $0.6$, and $0$. The $K_1\text{-}r_1$ and $K_1\text{-}r_2$ curves are identical only for $\beta_1=\beta_2$; otherwise they differ. Throughout all panels, the stability of the fixed points changes via saddle-node (SN) and pitchfork (PB) bifurcations, leading to multiple coexisting stable and unstable branches. For $\beta_1=1.3$ [Figs. \ref{fig1}(a) and \ref{fig1}(e)], phase-lag symmetry yields identical behavior in both layers, with a first-order explosive transition. The trivial fixed point $\mathcal{S}_0=(0,0)$ loses stability via a subcritical pitchfork bifurcation (PB1) at $K_1=7.47667$. From this point, two unstable branches emerge: one corresponds to the symmetric nontrivial state $\mathcal{S}_3=(r_1^{*},r_2^{*})$, which undergoes a saddle-node bifurcation (SN1) at $K_1=4.90146$ and continues as a stable strongly coherent branch; the other corresponds to the asymmetric state $\mathcal{S}_1=(r_1^{+},0)$ and remains unstable throughout. Owing to the symmetry between the layers, an analogous dynamics is observed for layer 2. For $\beta_1=1$ [Figs. \ref{fig1}(b) and \ref{fig1}(f)] compared to the previous case, layer 1 exhibits a double forward jump instead of a single jump, while layer 2 retains a first-order transition with narrower hysteresis. In layer 1, PB1 shifts to a lower value of $K_1~(K_1=3.70163)$, and only the asymmetric unstable branch $\mathcal{S}_1=(r_1^{+},0)$ emerges from PB1, which stabilizes via saddle-node bifurcation (SN2) at $K_1=3.60422$ and loses stability via subcritical pitchfork bifurcation (PB2) at $K_1=4.16104$. The nontrivial branch $\mathcal{S}_3=(r_1^{*},r_2^{*})$ emerges from PB2 and stabilizes via SN1 at $K_1=3.1822$, yielding a regime of weak synchronization for $K_1\in(3.60422,4.16104)$. In layer 2, the symmetric nontrivial steady state $\mathcal{S}_3=(r_1^{*},r_2^{*})$ corresponds to the synchronized branch, with PB1 and SN1 shifting to lower $K_1$ values compared to the $\beta_1=1.3$ case, resulting in a reduced hysteresis width. For $\beta_1=0.6$ [Fig. \ref{fig1}(c) and \ref{fig1}(g)], compared to the $\beta_1=1$ case, layer 1 exhibits a double jump in both forward and backward continuations: from incoherent to weak coherent and then to strong coherent in the forward direction, and the reverse in the backward direction. This results in two consecutive explosive transitions, each with its own hysteresis loop. The bifurcation points shift to PB1 at $K_1=2.42326$, SN2 at $K_1=1.96169$, PB2 at $K_1=3.32571$, and SN1 at $K_1=2.70824$. Bistability in $K_1\in(1.96169,2.42326)$ and $K_1\in(2.70824,3.32571)$ gives rise to two consecutive explosive jumps. Layer 2 continues to exhibit a first-order transition with further reduced hysteresis, with PB1 and SN1 shifting to even lower $K_1$ values. For $\beta_1=0$ [Figs. \ref{fig1}(d) and \ref{fig1}(h)], the bifurcation scenarios remain qualitatively similar to the $\beta_1=0.6$ case, with the bifurcation points shifting further to lower $K_1$ values. As $\beta_1$ decreases from $1.3$ to $0$, layer 1 transitions from a single explosive jump to double explosive jump, whereas layer 2 consistently exhibits a first-order transition with progressively narrower hysteresis. Thus, increasing the phase-lag asymmetry enhances the complexity of synchronization pathways in layer 1 while progressively weakening the hysteresis in layer 2, demonstrating that phase-lag asymmetry governs multistability and the emergence of double explosive transitions in the coupled system.

\begin{figure}
	\centering
	\includegraphics[width=6cm]{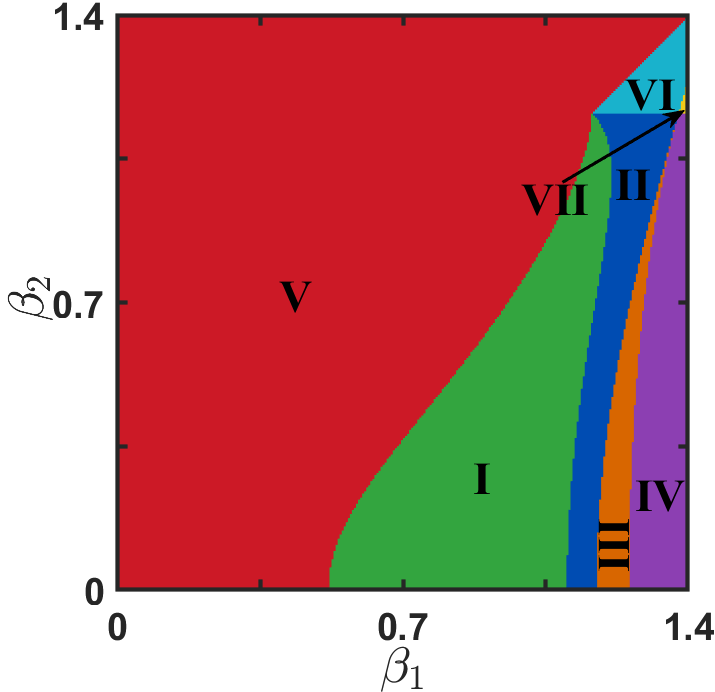}
	\caption{Phase diagram in the $\beta_1\text{-}\beta_2$ plane classifying the $K_1$-induced transitions of the second-layer order parameter $r_2$, for $K_{2}=5$, $p_1=p_2=2$, $h_1=h_2=2$, and $q=2$. The same seven regions (I--VII, colored and labeled as in Fig.~\ref{fig2}(a)) are recovered. The diagram is the mirror image of the $r_1$ phase diagram [Fig.~\ref{fig2}(a)] reflected across the line $\beta_1=\beta_2$: a transition type found at $(\beta_1,\beta_2)$ for $r_1$ occurs at the swapped point $(\beta_2,\beta_1)$ for $r_2$.}
	\label{fig3}
\end{figure}

\begin{figure*}
	\centering
	\includegraphics[width=15.5cm]{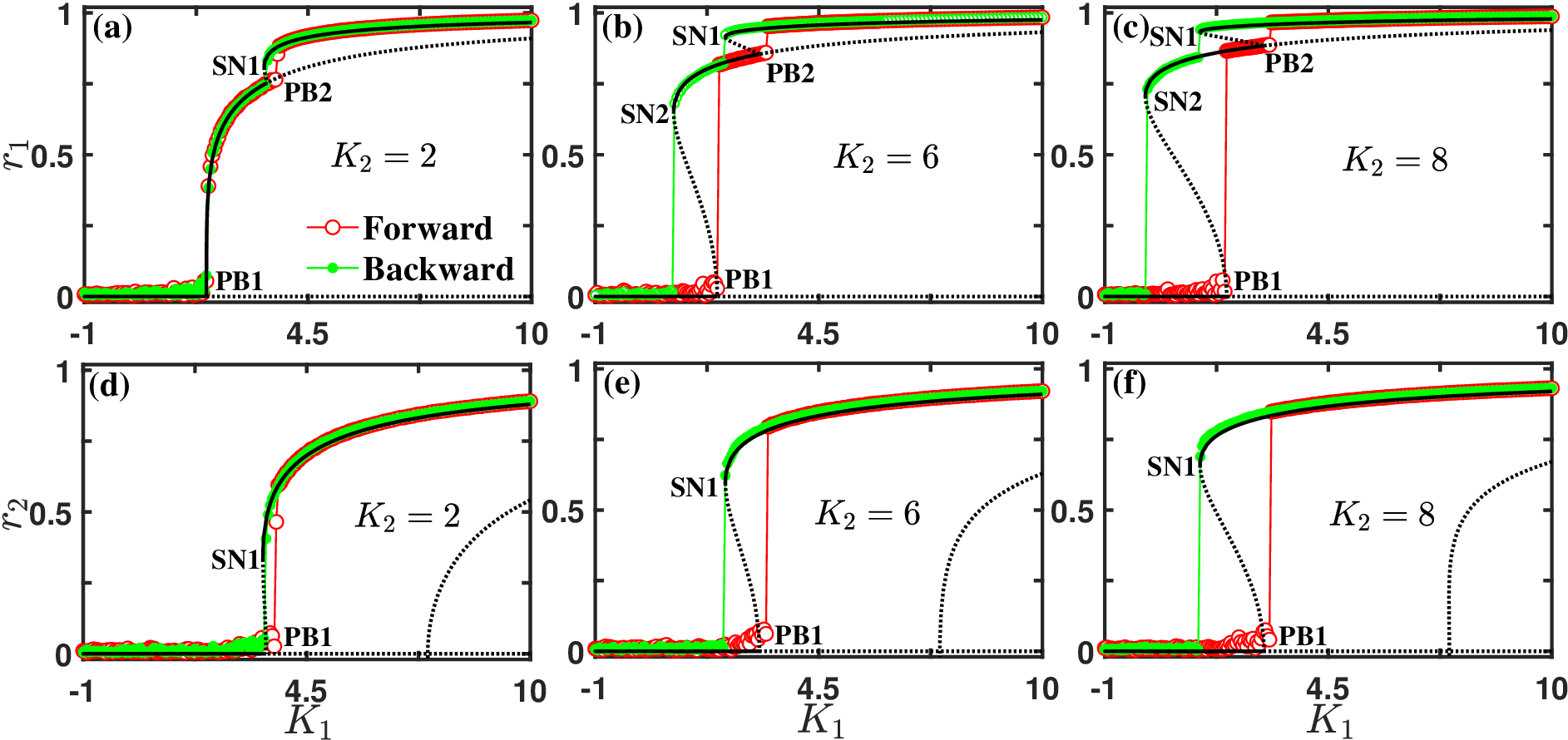}
	\caption{Evolution of the synchronization transitions with increasing $K_2$ at fixed $\beta_1=0$, $\beta_2=1.3$, $p_1=p_2=2$, $h_1=h_2=2$, and $q=2$. Top row: $r_1$ vs. $K_1$ for (a) $K_2=2$, (b) $K_2=6$, (c) $K_2=8$. As $K_2$ increases, the first layer undergoes a tiered transition of type II (a), followed by forward and backward double jumps with two distinct hysteresis loops (b), and finally forward and backward double jumps with a single hysteresis loop (c). Bottom row: $r_2$ vs. $K_1$ for the same $K_2$ values (d)-(f); the second layer stays classically explosive throughout, with the hysteresis width growing from (d) to (e) to (f) as $K_2$ increases. Forward (red, open circles) and backward (green, filled circles) branches are obtained from adiabatic direct numerical simulation of system~(\ref{eqn1}); black solid and dotted curves are the stable and unstable branches from the reduced-order model.}
	\label{fig4}
\end{figure*}


We now extend the analysis of Fig.~\ref{fig1} by systematically varying both $\beta_1$ and $\beta_2$ to map the transition scenarios in layer 1. Figure \ref{fig2} presents a comprehensive classification of the $K_1$-induced transition scenarios in layer 1 across the $\beta_1\text{-}\beta_2$ plane, for fixed $K_2=5$, $p_1=p_2=2$, $h_1=h_2=2$, and $q=2$. The phase diagram in Fig.~\ref{fig2}(a) reveals seven distinct regimes, each corresponding to a different synchronization route, as identified from the reduced model (\ref{eqn11}) and validated against the microscopic simulations. In Regime I (forest green), the system exhibits multiple transitions that depend on the initial conditions [Fig.~\ref{fig2}(b)], which is in agreement with previous findings~\cite{das2026effect}. The strongly coherent and incoherent branches are connected to a weakly synchronized middle branch via two unstable paths. In the forward adiabatic continuation with random initial phases (red open circles), the system jumps from incoherence to the stable strongly coherent branch, while in the backward continuation (green filled circles), it jumps directly to the incoherent state, yielding a hysteresis loop. The middle branch, however, can be accessed by choosing initial conditions with one layer nearly synchronized and the other randomly distributed (light blue open circles), revealing an additional forward route from incoherence to weak synchronization and subsequently to strong synchronization. Regime II (navy blue) is characterized by two forward jumps but only a single backward jump to the incoherent state, whereas Regime III (burnt orange) exhibits two jumps in both directions with a single hysteresis structure [Figs.~\ref{fig2}(c) and \ref{fig2}(d)]. Regime IV (violet) displays two jumps in both directions with two well-defined hysteresis loops [Fig.~\ref{fig2}(e)]. Regime V (crimson red) represents the classical explosive transition to synchronization, as confirmed by Figs.~\ref{fig2}(f) and \ref{fig2}(g). Finally, Regimes VI (teal) and VII (gold) correspond to two types of tiered synchronization~\cite{skardal2022tiered,biswas2024effect}. In type I, the system transitions continuously from incoherence to weak synchronization and then jumps abruptly to strong synchronization, with a direct backward jump to incoherence. In type II, the system jumps from the strongly synchronized state to the weakly synchronized state during the backward transition. These are further confirmed by Figs.~\ref{fig2}(h) and \ref{fig2}(i).

\begin{figure*}
	\centering
	\includegraphics[width=16cm]{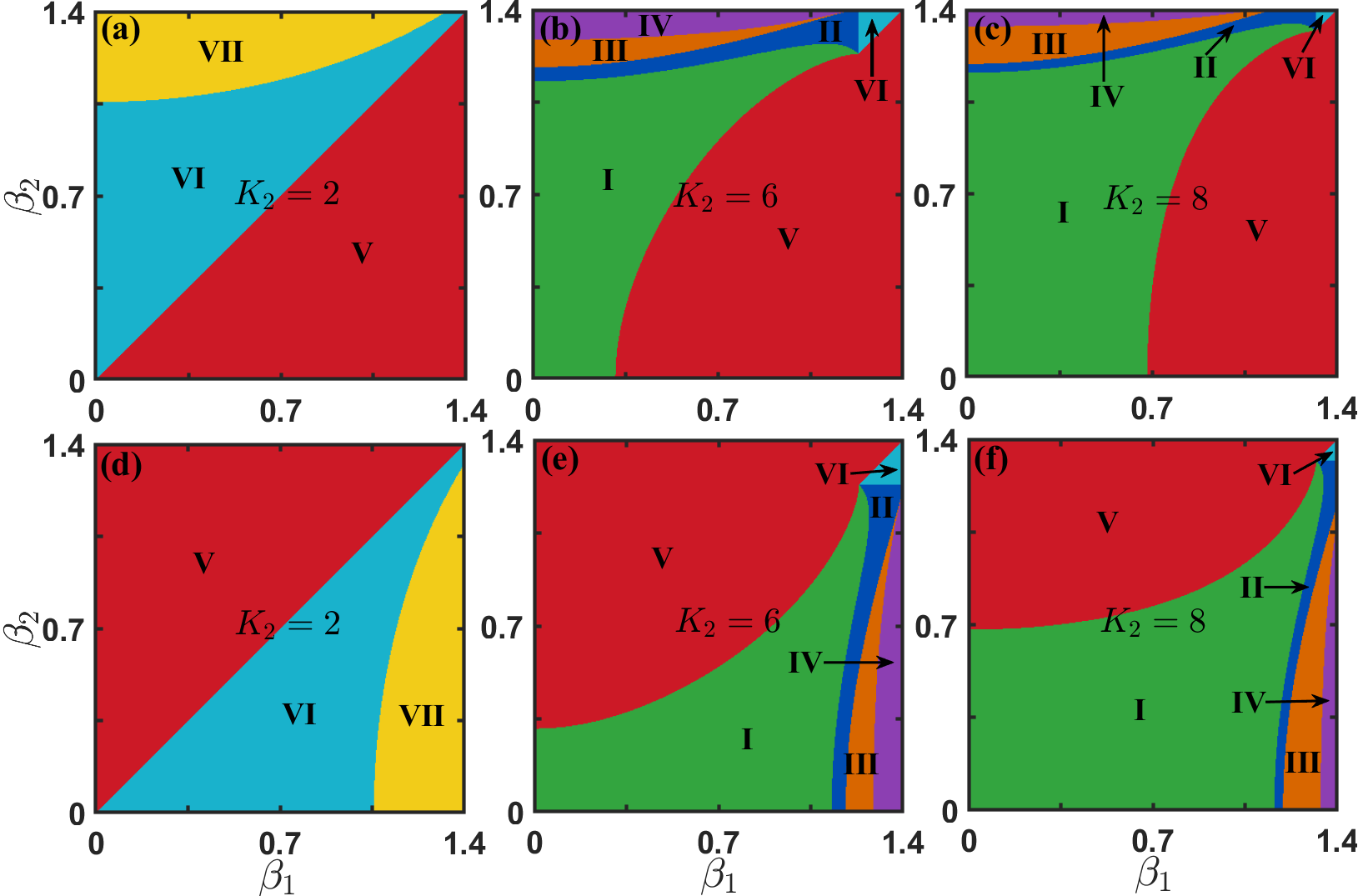}
	\caption{Effect of $K_2$ on the transition regimes classified in the $\beta_1\text{-}\beta_2$ plane, obtained from the reduced-order model by varying $K_1$, for $p_1=p_2=2$, $h_1=h_2=2$, and $q=2$; Roman numerals label the regimes defined in Fig.~\ref{fig2}(a). Top row: $r_1$ phase diagrams for (a) $K_2=2$, (b) $K_2=6$, and (c) $K_2=8$. At $K_2=2$ (a), only regions V, VI, and VII appear. At $K_2=6$ (b), region VII vanishes, region VI shrinks, and regions I--VI are present. At $K_2=8$ (c), the same six regions remain present, with regions II, IV, V, and VI shrinking and regions I and III expanding relative to (b). Bottom row: $r_2$ phase diagrams for the corresponding $K_2$ values in (d)-(f) exhibit the same reorganization and are mirror images of (a)-(c) under reflection across the line $\beta_1=\beta_2$: a transition type occurring at $(\beta_1,\beta_2)$ for $r_1$ occurs at $(\beta_2,\beta_1)$ for $r_2$.}
	\label{fig5}
\end{figure*}

The boundaries between these regimes are determined by the existence and relative ordering of the bifurcation points $K_1^{\mathrm{SN1}}$, $K_1^{\mathrm{SN2}}$, $K_1^{\mathrm{PB1}}$, and $K_1^{\mathrm{PB2}}$, which can be derived analytically from the stability analysis of the fixed points. The pitchfork bifurcation $K_1^{\mathrm{PB1}}$, where the trivial state $\mathcal{S}_0=(0,0)$ loses stability, is given by
\begin{equation}
	K_1^{\mathrm{PB1}}=\min\left(\frac{2}{\cos\beta_1}, \frac{2}{\cos\beta_2} \right). \label{eqn14}
\end{equation}
The saddle-node bifurcation $K_1^{\mathrm{SN2}}$, marking the birth of the stable asymmetric single-layer state $\mathcal{S}_1=(r_1^{+},0)$, is obtained as
\begin{equation}
	K_1^{\mathrm{SN2}}=\frac{2}{(1-{r_1^{+}}^2)\cos\beta_1}-K_2{r_1^{+}}^2, \label{eqn15}
\end{equation}
with $r_1^{+}=\sqrt{1-\sqrt{\frac{2}{K_2\cos\beta_1}}}$ and $r_1^{+}\in(0,1]$, provided the stability condition $\max\left(J_{11}^{(1)}, J_{22}^{(1)}\right)<0$ is satisfied. The pitchfork bifurcation $K_1^{\mathrm{PB2}}$, where the weak coherent state $\mathcal{S}_1=(r_1^{+},0)$ loses stability, is determined by solving
\begin{equation}\label{eqn16}
	\begin{aligned}
		&K_2q{r_1^{+}}^{p+4}\cos\beta_1 \cos\beta_2-K_2q{r_1^{+}}^{p+2}\cos\beta_1\cos\beta_2 \\
		&+K_2{r_1^{+}}^4\cos\beta_1\cos\beta_2-K_2{r_1^{+}}^2\cos\beta_1\cos\beta_2 \\
		&+2q{r_1^{+}}^{p}\cos\beta_2+2{r_1^{+}}^{2}\cos\beta_1+2\cos\beta_2-2\cos\beta_1=0
	\end{aligned}
\end{equation}
with $r_1^{+}\in(0,1]$, and then computing
\begin{equation}
	K_1^{\mathrm{PB2}}=\frac{2}{(1+q{r_1^{+}}^{p})\cos\beta_2}, \label{eqn17}
\end{equation}
subject to the stability condition $\max\left(J_{11}^{(1)}, J_{22}^{(1)}\right)<0$. Finally, the saddle-node bifurcation $K_1^{\mathrm{SN1}}$, where the fully coherent nontrivial stable state $\mathcal{S}_3=(r_1^{*},r_2^{*})$ emerges, corresponds to the minimum value of $K_1$ for which the two stationary-state equations
\begin{subequations}\label{eqn18}
	\begin{equation}
		K_1=\frac{2/(1-{r_1^{*}}^2)\cos\beta_1-K_2(1+q{r_2^{*}}^{h}){r_1^{*}}^2}{(1+q{r_2^{*}}^{p})}, \label{eqn18(a)}
	\end{equation}
	\begin{equation}
		K_1=\frac{2/(1-{r_2^{*}}^2)\cos\beta_2-K_2(1+q{r_1^{*}}^{h}){r_2^{*}}^2}{(1+q{r_1^{*}}^{p})}, \label{eqn18(b)}
	\end{equation}
\end{subequations}
are simultaneously satisfied; this minimum is determined numerically.

The relative ordering of these bifurcation points dictates the transition scenario. When all four bifurcation points exist, the following scenarios arise: 
\begin{enumerate}[(i)]
	\item If $K_1^{\mathrm{SN1}}<K_1^{\mathrm{SN2}}<K_1^{\mathrm{PB1}}<K_1^{\mathrm{PB2}}$, the system exhibits a double forward jump and a single backward jump
	
	\item If $K_1^{\mathrm{SN2}}<K_1^{\mathrm{SN1}}<K_1^{\mathrm{PB1}}<K_1^{\mathrm{PB2}}$, the system exhibits double jumps in both the forward and backward directions with a single hysteresis structure
	
	\item If $K_1^{\mathrm{SN2}}<K_1^{\mathrm{PB1}}<K_1^{\mathrm{SN1}}<K_1^{\mathrm{PB2}}$, the system exhibits double jumps in both the forward and backward directions with two distinct hysteresis loops
	
	\item If $K_1^{\mathrm{SN1}}<K_1^{\mathrm{SN2}}<K_1^{\mathrm{PB2}}<K_1^{\mathrm{PB1}}$, the system shows initial-condition-dependent multiple transitions
	
	\item If $K_1^{\mathrm{SN2}}<K_1^{\mathrm{SN1}}<K_1^{\mathrm{PB2}}<K_1^{\mathrm{PB1}}$, the system exhibits two backward jumps and a single forward jump
	
\end{enumerate}
When PB1 and PB2 are both supercritical, the saddle-node bifurcations SN1 and SN2 do not exist. In this case, if $K_1^{\mathrm{PB1}}<K_1^{\mathrm{PB2}}$, the system undergoes a continuous transition to synchronization. When SN2 and PB2 are absent, the system reduces to the classical explosive transition if $K_1^{\mathrm{SN1}}<K_1^{\mathrm{PB1}}$. If PB1 is supercritical while PB2 is subcritical, SN2 is absent, and two tiered transition scenarios emerge: if $K_1^{\mathrm{SN1}}<K_1^{\mathrm{PB1}}<K_1^{\mathrm{PB2}}$, the system exhibits tiered transition of type I; if $K_1^{\mathrm{PB1}}<K_1^{\mathrm{SN1}}<K_1^{\mathrm{PB2}}$, the system exhibits tiered transition of type II.

To validate these analytical expressions, we select a representative point $(\beta_1,\beta_2)=(0.1,1.3)$ from Regime IV in the phase diagram of Fig.~\ref{fig2}(a). The computed bifurcation points are $K_1^{\mathrm{PB1}}=2.01$, $K_1^{\mathrm{SN2}}=1.96169$, $K_1^{\mathrm{PB2}}=3.1265$, and $K_1^{\mathrm{SN1}}=1.3246$. Their relative ordering, $K_1^{\mathrm{SN2}}<K_1^{\mathrm{PB1}}<K_1^{\mathrm{SN1}}<K_1^{\mathrm{PB2}}$, confirms the occurrence of double explosive transitions in both forward and backward direction, in full agreement with the numerical results presented in Fig.~\ref{fig2}(e). This demonstrates excellent agreement between the analytical expressions and the numerical continuation, thereby validating the derived bifurcation conditions and the phase diagram classification.

Having characterized the transition scenarios in layer 1, we now investigate the corresponding behavior of layer 2. A natural question arises: for which values of $(\beta_1,\beta_2)$ does layer 2 exhibit double explosive transitions? To address this, we compute the steady-state order parameter $r_2$ from the reduced model (\ref{eqn11}) as a function of $K_1$ for the same parameter set, and construct a surface plot of $r_2$ in the $\beta_1\text{-}\beta_2$ plane, as shown in Fig.~\ref{fig3}. The resulting phase diagram, presented in Fig.~\ref{fig3}, recovers the same seven regimes (I--VII) as in Fig.~\ref{fig2}(a), with the colors and labels retained for consistency. However, the diagram for $r_2$ is the mirror image of the $r_1$ phase diagram reflected across the line $\beta_1=\beta_2$. This symmetry arises from the interchangeability of the layers in the reduced model (\ref{eqn11}): a transition scenario observed for layer 1 at a given parameter pair $(\beta_1,\beta_2)$ appears for layer 2 at the swapped pair $(\beta_2,\beta_1)$. Thus, the phase diagram for $r_2$ mirrors that of $r_1$, confirming the symmetric structure of the bifurcation scenarios in the two-layer system.

Having established the phase diagram for fixed $K_2=5$, we now examine the role of $K_2$ on the transition scenarios in both layers. Figure~\ref{fig4} presents the bifurcation diagrams of $K_1$ vs. $r_1$ [Figs.~\ref{fig4}(a)--\ref{fig4}(c)] and $r_2$ [Figs.~\ref{fig4}(d)--\ref{fig4}(f)] for $K_2=2,6,8$ with $\beta_1=0$, $\beta_2=1.3$, $p_1=p_2=2$, $h_1=h_2=2$, and $q=2$. For $K_2=2$ [Figs.~\ref{fig4}(a) and \ref{fig4}(d)], layer 1 exhibits a tiered transition of type II: a continuous forward path from incoherence to weak coherence, then an abrupt jump to strong coherence, with the reverse in the backward direction. The incoherent state loses stability via a supercritical pitchfork bifurcation (PB1) at $K_1=2$, giving rise to a stable branch that subsequently loses stability via a subcritical pitchfork bifurcation (PB2) at $K_1=3.49849$, with an unstable branch emerging from PB2 and stabilizing via saddle-node bifurcation (SN1) at $K_1=3.422$. In contrast, layer 2 exhibits a classical explosive transition with a narrow hysteresis loop, where the incoherent state loses stability via subcritical pitchfork bifurcation (PB1) at $K_1=3.49849$, and the unstable branch stabilizes via SN1 at $K_1=3.422$. Increasing $K_2$ to $6$ [Figs.~\ref{fig4}(b) and \ref{fig4}(e)] fundamentally alters the transition in layer 1 to forward and backward double jump transitions, with two distinct hysteresis loops. The incoherent state loses stability via a subcritical pitchfork bifurcation (PB1) at $K_1=2$, from which an unstable branch emerges, stabilizes via saddle-node bifurcation (SN2) at $K_1=0.928203$, and loses stability via subcritical pitchfork bifurcation (PB2) at $K_1=3.03807$. From PB2, another unstable branch emerges and stabilizes via saddle-node bifurcation (SN1) at $K_1=2.20096$. In layer 2, the classical explosive transition is preserved, but the hysteresis width increases significantly, with PB1 shifting to $K_1=3.03807$ and SN1 to $K_1=2.20096$. For $K_2=8$ [Figs.~\ref{fig4}(c) and \ref{fig4}(f)], layer 1 continues to exhibit forward and backward double jump transition, but with a single hysteresis structure. The bifurcation structure remains qualitatively similar to the $K_2=6$ case. In layer 2, the classical explosive transition persists with further widening of the hysteresis loop. Thus, increasing $K_2$ systematically alters the transitions: layer 1 evolves from a tiered transition to a forward and backward double jump transitions with two distinct hysteresis loops, and subsequently to one with a single hysteresis loop, while layer 2 consistently exhibits classical explosive transitions with progressively widening hysteresis.

We now provide a comprehensive view of the effect of $K_2$ on the synchronization transitions across the entire $\beta_1\text{-}\beta_2$ plane. Figure~\ref{fig5} presents surface plots of the steady-state order parameter $r_1$ [Figs.~\ref{fig5}(a)--\ref{fig5}(c)] and $r_2$ [Figs.~\ref{fig5}(d)--\ref{fig5}(f)] as functions of $K_1$ in the $\beta_1\text{-}\beta_2$ plane for $K_2=2,6,8$, with the same color and label conventions as in Fig.~\ref{fig2}(a). For $K_2=2$ [Fig.~\ref{fig5}(a)], only three regimes persist: Regimes V (classical explosive), VI (tiered type I), and VII (tiered type II). Increasing $K_2$ to $6$ [Fig.~\ref{fig5}(b)] expands the phase diagram to include Regimes I--V, while Regime VII disappears and Regime V shrinks significantly. For $K_2=8$ [Fig.~\ref{fig5}(c)], Regimes II, IV, V, and VI shrink, whereas Regimes I and III widen compared to the $K_2=6$ case. The surface plots for $r_2$ [Figs.~\ref{fig5}(d)--\ref{fig5}(f)] exhibit behavior analogous to that of $r_1$, with the same regimes appearing for each corresponding value of $K_2$. As observed in Fig.~\ref{fig3}, the $r_2$ diagrams are mirror images of the $r_1$ diagrams reflected across the line $\beta_1=\beta_2$, consistent with the layer symmetry of the model. From this point onward, we restrict our analysis to layer 1, as the behavior of layer 2 follows from the layer symmetry established in Figs.~\ref{fig3} and \ref{fig5}.

\begin{figure}
    \centering
    \includegraphics[width=\linewidth]{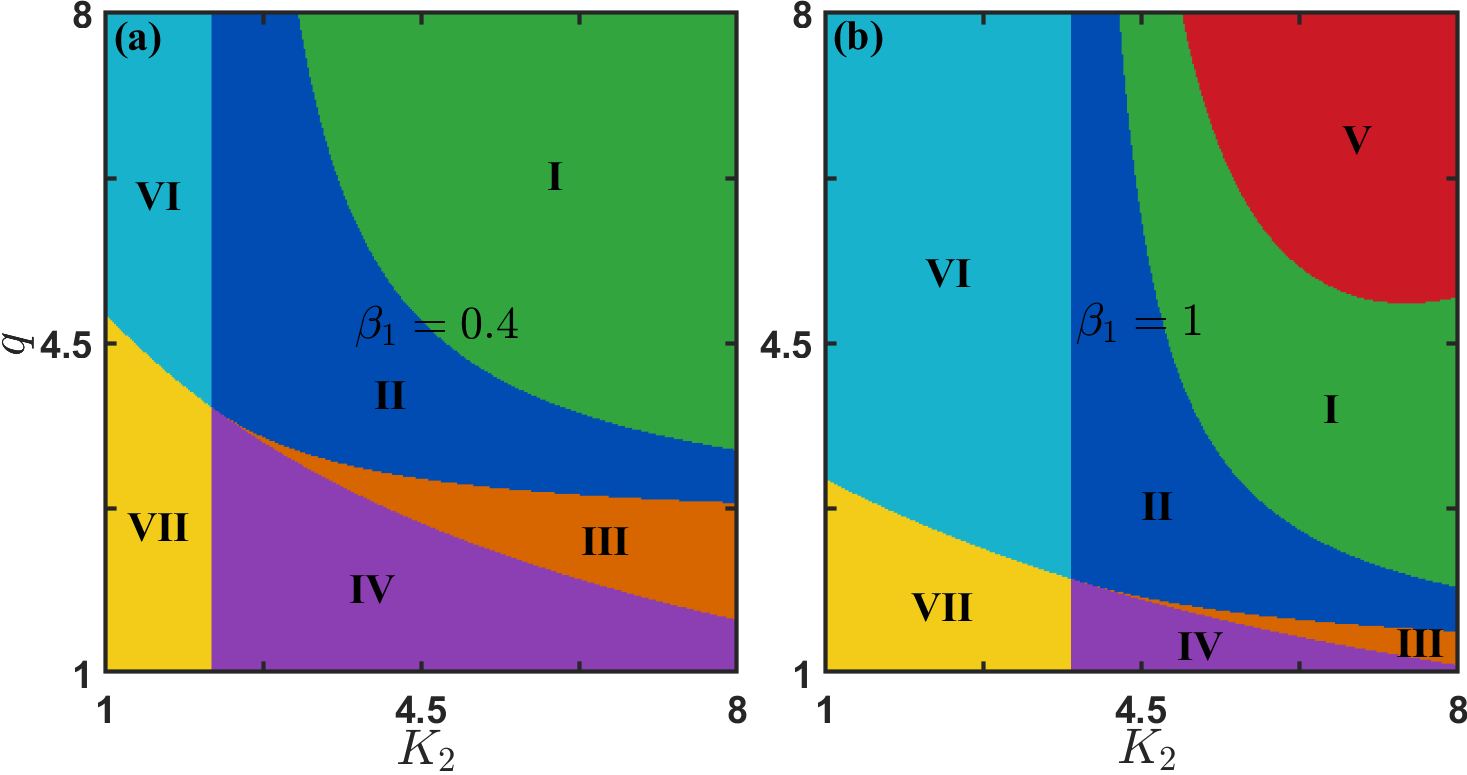}
    \caption{Combined effect of $K_2$, $q$, and $\beta_1$ on the transition regime of $r_1$, for $p_1=p_2=2$, $h_1=h_2=2$, and $\beta_2=1.3$. Phase diagram in the $K_2\text{-}q$ plane, classified from $K_1$ sweeps in the reduced-order model, for (a) $\beta_1=0.4$ and (b) $\beta_1=1$; roman numerals label the same regimes as in Fig.~\ref{fig2}(a). (a) At $\beta_1=0.4$, six regions appear: I, II, III, IV, VI, VII. (b) At $\beta_1=1$, all six regions persist and region V (classical explosive synchronization) additionally appears.}
    \label{fig6}
\end{figure}

\begin{figure*}
	\centering
	\includegraphics[width=\linewidth]{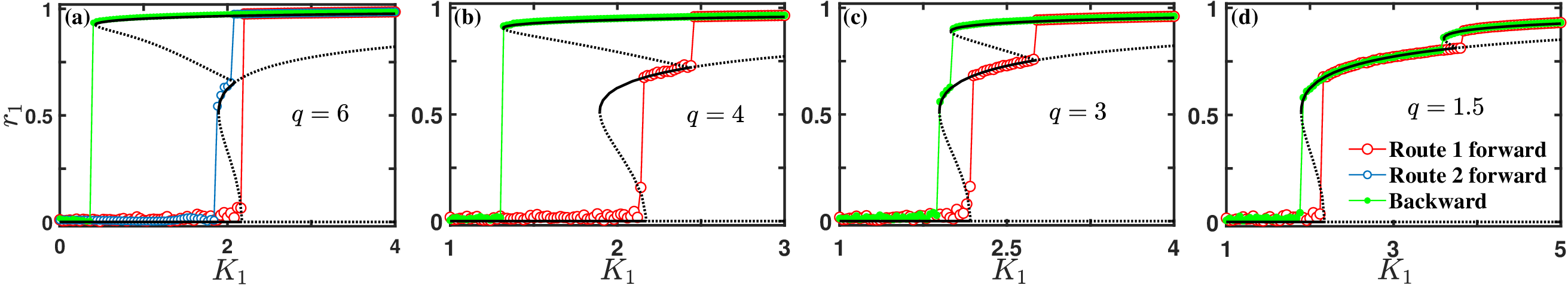}
	\caption{Effect of $q$ on the synchronization transition, for $K_2=4$,$p_1=p_2=2$, $h_1=h_2=2$, $\beta_1=0.4$, and $\beta_2=1.3$. $r_1$ vs. $K_1$ for (a) $q=6$, (b) $q=4$, (c) $q=3$, (d) $q=1.5$. As $q$ decreases, the transition scenario changes from multiple transitions dependent on the initial condition (a), to a double jump on the forward branch with a single jump on the backward branch (b), to a double jump on both branches with a single hysteresis loop (c), to a double jump on both branches with two distinct hysteresis loops (d). Black solid and dotted curves are the stable and unstable branches from the reduced-order model; red (open circles) and green (filled circles) curves are the forward and backward adiabatic branches from direct numerical simulation of system~(\ref{eqn1}); the additional forward route reached from a different initial condition in (a) is shown in light blue (open circles).}
	\label{fig7}
\end{figure*}

Having established the role of $K_2$ on the transition scenarios across the $\beta_1\text{-}\beta_2$ plane, we now turn to the combined effects of $K_2$ and and the adaptation strength $q$. How do these two parameters jointly influence the synchronization transitions in layer 1? To address this, we construct surface plots of the transition regimes in the $K_2\text{-}q$ plane for two values of $\beta_1=0.4$ and $1$, with fixed $p=2$, $h=2$, and $\beta_2=1.3$. The color and label conventions follow those of Fig.~\ref{fig2}(a). For $\beta_1=0.4$ [Fig.~\ref{fig6}(a)], six regimes appear: I, II, III, IV, VI, and VII. Varying $q$ for a fixed $K_2$ systematically alters the transition type. Decreasing $q$ from $8$ to $1$ transforms Regime VI to VII for certain $K_2$ values, while for other $K_2$ values, it leads to a sequence I $\rightarrow$ II $\rightarrow$ III $\rightarrow$ IV. Increasing $\beta_1$ to $1$ [Fig.~\ref{fig6}(b)] introduces Regime V while retaining the other six regimes. Compared to $\beta_1=0.4$, Regime VI expands in both directions, Regime VII narrows in $q$ but widens in $K_2$, Regime II narrows in $K_2$ but widens in $q$, and Regimes I, III, and IV shrink.

To further elucidate the role of $q$, we select four representative points $(K_2,q)=(4,6),(4,4),(4,3),(4,1.5)$ from the phase diagram in Fig.~\ref{fig6}(a). The corresponding bifurcation diagrams are shown in Figs.~\ref{fig7}(a)--\ref{fig7}(d), with both analytical (reduced model) and numerical (full system) results in excellent agreement. For $q=6$ [Fig.~\ref{fig7}(a)], the system exhibits multiple transitions dependent on initial conditions. Reducing $q$ to $4$ [Fig.~\ref{fig7}(b)] yields a double forward jump and a single backward jump, with the weak synchronization branch beginning to widen compared to the $q=6$ case. For $q=3$ [Fig.~\ref{fig7}(c)], the system exhibits a forward and backward double jump transition with single hysteresis loop, and the weak synchronization branch widens further. At $q=1.5$ [Fig.~\ref{fig7}(d)], forward and backward double jump synchronization transition emerges, but with two distinct hysteresis loops, and the weak synchronization branch widens even more. The sequence of transitions observed for decreasing $q$ directly follows the regimes identified in the $K_2\text{-}q$ phase diagram.

\begin{figure}
    \centering
    \includegraphics[width=\linewidth]{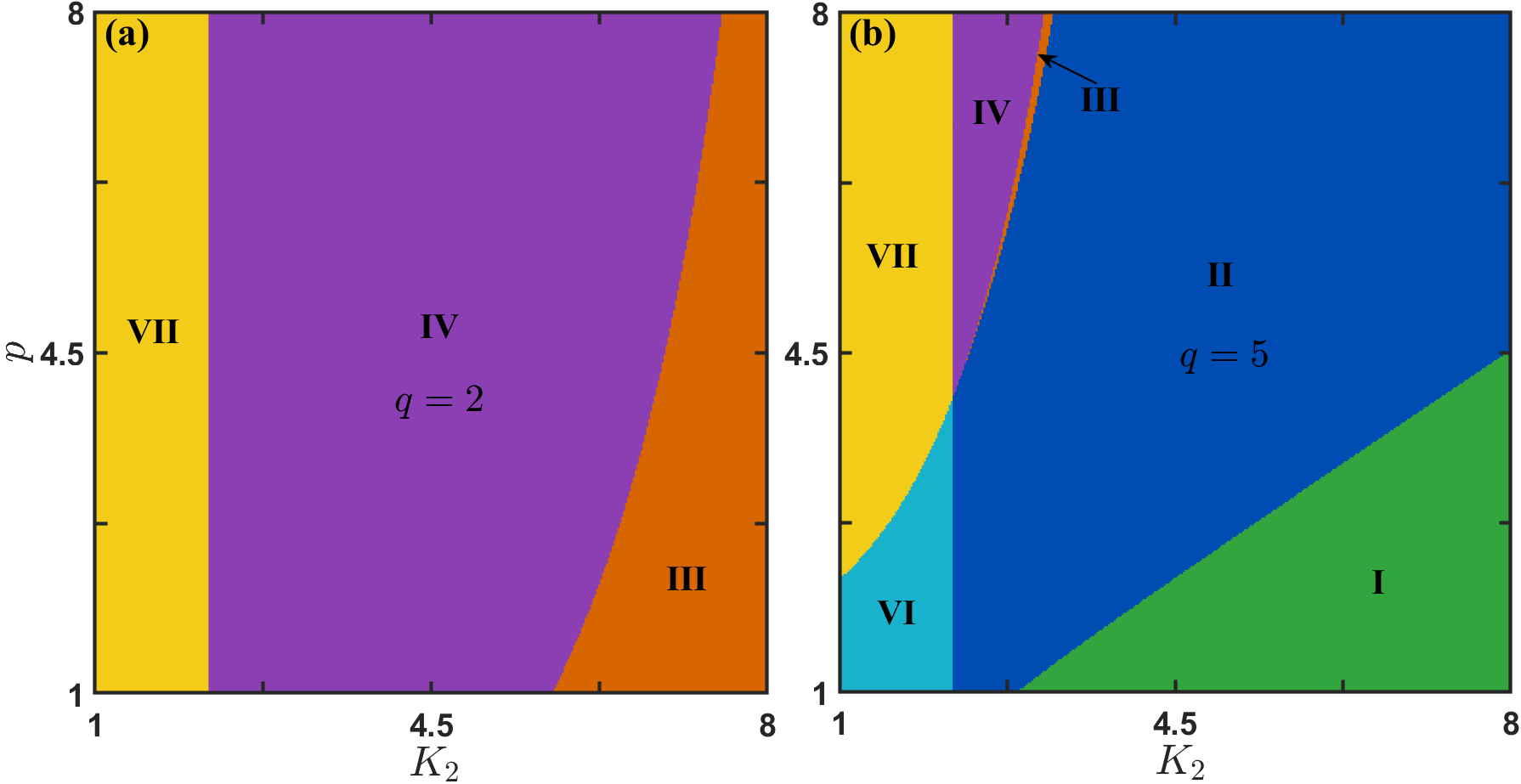}
    \caption{Combined effect of $K_2$, $p$, and $q$ on the transition regimes of $r_1$, for $h=2$, $\beta_1=0.4$, and $\beta_2=1.3$. Phase diagram in the $K_2\text{-}p$ plane, classified from $K_1$ sweeps in the reduced-order model; roman numerals and colors denote the same regimes as in Fig.~\ref{fig2}(a). (a) At $q=2$, only three regions appear: III, IV, and VII. (b) Increasing $q$ to $5$ shrinks all three regions relative to (a) and allows three further regions---I, II, and VI---to emerge. Thus, raising $q$ diversifies the transition regimes accessible in the $K_2\text{-}p$ plane.}
    \label{fig8}
\end{figure}

What role does the adaptation exponent $p$ play in shaping the synchronization transitions? Having explored the influence of $q$, we now turn our attention to $p$, which governs the nonlinearity of the pairwise feedback. Figure~\ref{fig8} presents surface plots of the transition regimes in layer 1 in the $K_2\text{-}p$ plane for two values of $q=2$ and $5$, with fixed $h=2$, $\beta_1=0.4$, and $\beta_2=1$. For $q=2$ [Fig.~\ref{fig8}(a)], only three regimes appear: III, IV, and VII. For a fixed $p$, increasing $K_2$ from $1$ to $8$ drives the system through a sequence of regimes VII $\rightarrow$ IV $\rightarrow$ III. However, for $K_2<5.76$, varying $p$ has no discernible effect on the transition type. Beyond this threshold, increasing $p$ transforms Regime III to IV. Increasing $q$ to $5$ [Fig.~\ref{fig8}(b)] introduces Regimes I, II, and VI, while Regimes III, IV, and VII shrink. In this case, increasing $p$ for fixed $K_2$ yields several distinct sequences: VI $\rightarrow$ VII, II $\rightarrow$ IV, II $\rightarrow$ III $\rightarrow$ IV, and I $\rightarrow$ II, depending on the value of $K_2$.

\begin{figure}
    \centering
    \includegraphics[width=\linewidth]{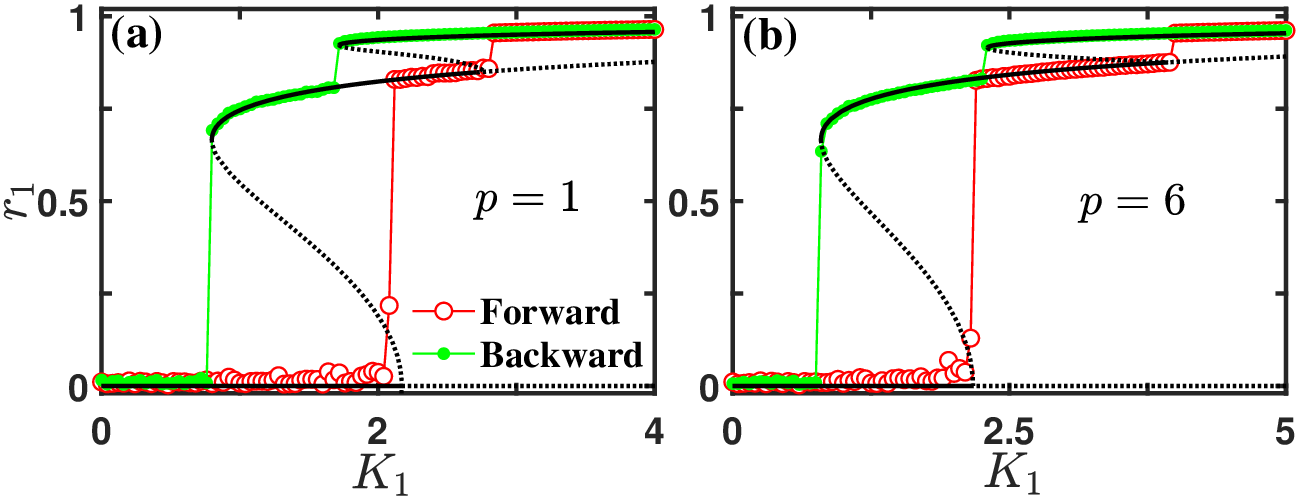}
    \caption{Role of $p$ in shaping the synchronization transition of $r_1$, for $K_2=7$, $q=2$, $h=2$, $\beta_1=0.4$, and $\beta_2=1.3$. Bifurcation diagrams of $r_1$ vs. $K_1$ are shown for (a) $p=1$ and (b) $p=6$. Increasing $p$ transforms the forward and backward double jump transition from a single hysteresis loop at $p=1$ (a) into two distinct hysteresis loops at $p=6$ (b). Black solid and dotted curves show stable and unstable branches of the reduced-order model; red (open circles) and green (filled circles) curves show forward and backward adiabatic continuations from direct numerical simulations of Eq.~(\ref{eqn1}).}
    \label{fig9}
\end{figure}

To further illustrate the role of $p$, we select two representative points $(K_2,p)=(7,1)$ and $(7,6)$ from Fig.~\ref{fig8}(a). The corresponding bifurcation diagrams are shown in Figs.~\ref{fig9}(a) and \ref{fig9}(b). For $p=1$ [Fig.~\ref{fig9}(a)], the system exhibits forward and backward double jump transitions with a single hysteresis loop, whereas increasing $p$ to $6$ [Fig.~\ref{fig9}(b)] separates the two explosive transitions, resulting in two distinct hysteresis loops in both directions, with the weak synchronization branch widening significantly compared to the $p=1$ case.

\begin{figure}
    \centering
    \includegraphics[width=\linewidth]{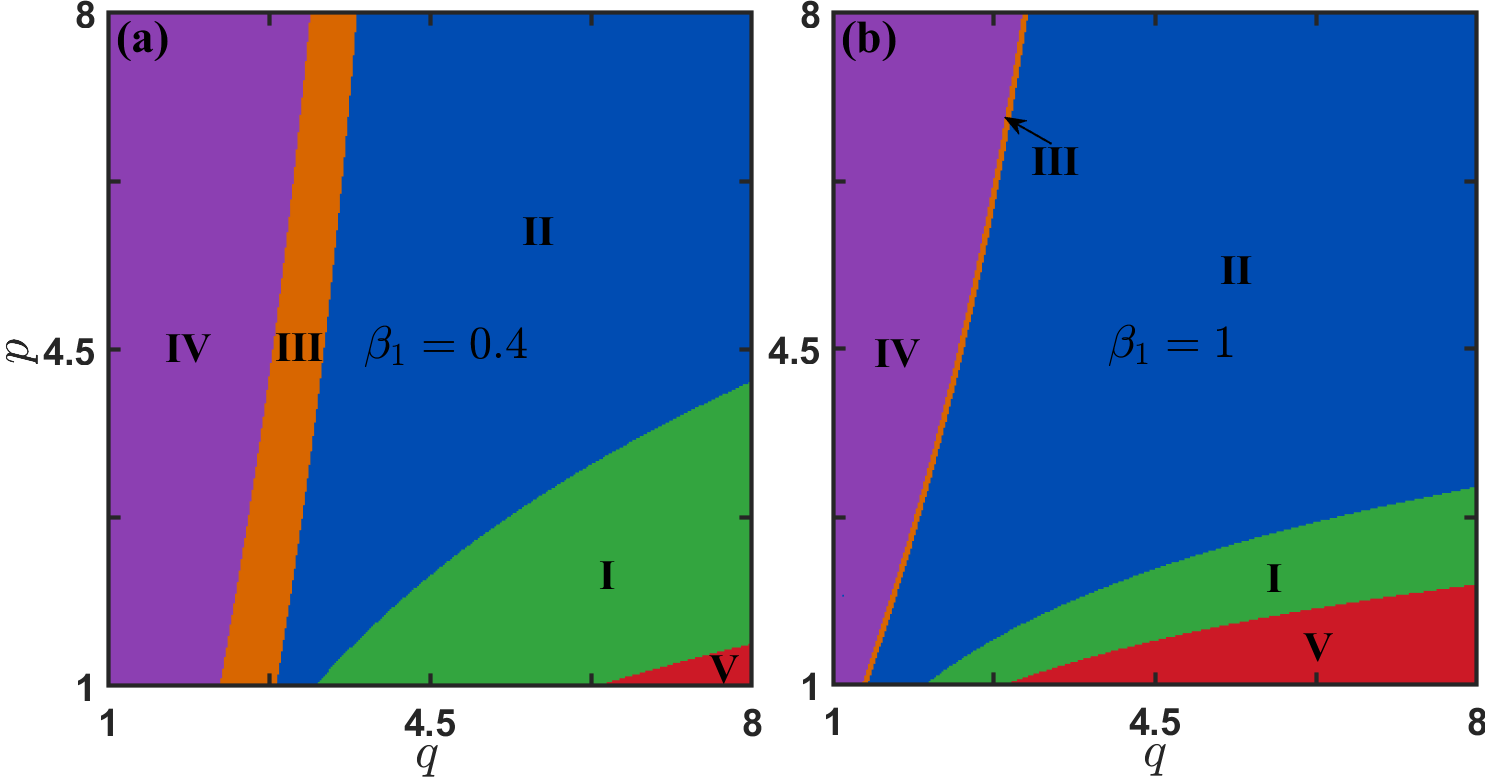}
    \caption{Combined effect of $q$, $p$, and $\beta_1$ on the transition regimes of $r_1$, classified from $K_1$ sweeps of the reduced-order model, for $K_{2}=5$, $h=2$, and $\beta_2=1.3$; roman numerals and colors follow the classification of Fig.~\ref{fig2}(a). (a) At $\beta_1=0.4$, all five regimes I--V coexist in the $q\text{-}p$ plane. (b) Increasing $\beta_1$ to $1$ preserves the same five regimes but redistributes their extent: regions I, III, and IV shrink, while regions II and V expand correspondingly.}
    \label{fig10}
\end{figure}

We now turn to the combined effects of $q$ and $p$ on the transition scenarios. Figure~\ref{fig10} presents surface plots of the transition regimes in the $q\text{-}p$ plane for two values of $\beta_1=0.4$ and $1$, with fixed $K_{2}=5$, $h=2$, and $\beta_2=1.3$. For $\beta_1=0.4$ [Fig.~\ref{fig10}(a)], five regimes appear: I, II, III, IV, and V. For a fixed $p$, increasing $q$ drives the system through the sequence IV $\rightarrow$ III $\rightarrow$ II, with some $p$ values extending this to IV $\rightarrow$ III $\rightarrow$ II $\rightarrow$ I or even IV $\rightarrow$ III $\rightarrow$ II $\rightarrow$ I $\rightarrow$ V. For $q<2.22$, varying $p$ has no effect. Beyond this threshold, increasing $p$ induces transitions such as III $\rightarrow$ IV, II $\rightarrow$ III $\rightarrow$ IV, I $\rightarrow$ II, or V $\rightarrow$ I $\rightarrow$ II. Increasing $\beta_1$ to $1$ [Fig.~\ref{fig10}(b)] preserves the same set of regimes, but Regimes I, III, and IV shrink while Regimes II and V widen.

\begin{figure}
    \centering
    \includegraphics[width=\linewidth]{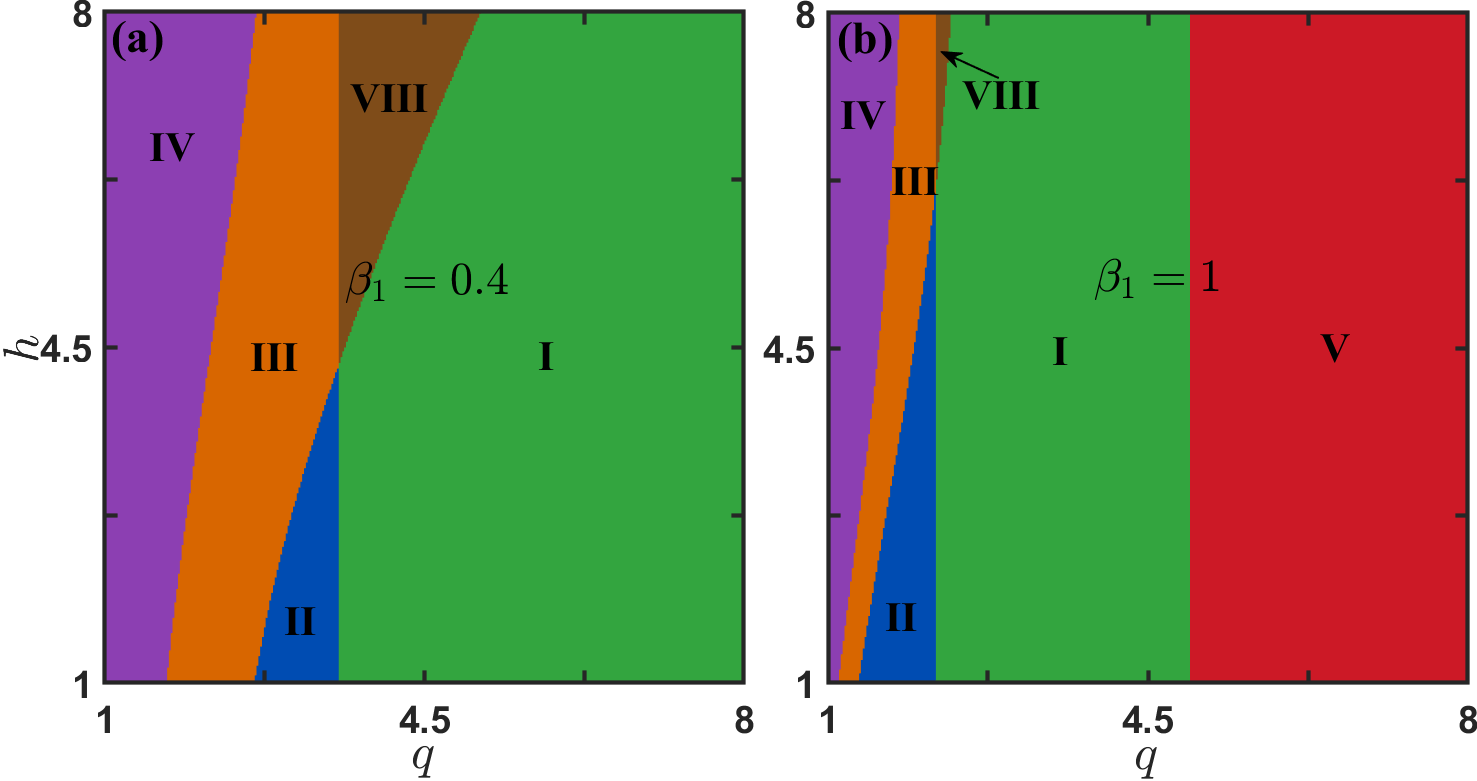}
    \caption{Combined effect of $q$, $h$, and $\beta_1$ on the transition regimes of $r_1$ in the $q\text{-}h$ plane, classified from $K_1$ sweeps of the reduced-order model, for $K_{2}=7$, $p=2$, and $\beta_2=1.3$. Regions I--IV retain the same roman numeral labels and colors as in Fig.~\ref{fig2}(a). The brown-shaded region VIII denotes a newly identified transition scenario absent in Fig.~\ref{fig2}, in which the forward continuation exhibits one jump whereas the backward continuation exhibits two, making it the mirror counterpart of region II. (a) At $\beta_1=0.4$, regions I--IV coexist with region VIII. (b) Increasing $\beta_1$ to $1$ shrinks regions I, III, IV, and VIII while expanding region II; region V, absent in (a), also emerges.}
    \label{fig11}
\end{figure}

The sequence of transitions observed for varying $q$ and $p$ naturally leads to the question: what role does the higher-order adaptation exponent $h$ play in shaping the synchronization scenarios? To address this, we now investigate the combined effects of $q$ and $h$ on the transition regimes in layer 1. Figure~\ref{fig11} presents surface plots of the transition regimes in the $q\text{-}h$ plane for two values of $\beta_1=0.4$ and $1$, with fixed $K_{2}=7$, $p=2$, and $\beta_2=1.3$. For $\beta_1=0.4$ [Fig.~\ref{fig11}(a)], four regimes appear: I, II, III, and IV, along with a new regime VIII (shaded brown), where the system exhibits a single forward jump from incoherence to strong coherence, but two backward jumps from strong to weak coherence and then to incoherence. For fixed $h$, increasing $q$ drives the system through sequences such as IV $\rightarrow$ III $\rightarrow$ II $\rightarrow$ I or IV $\rightarrow$ III $\rightarrow$ VIII $\rightarrow$ I, depending on the value of $h$. For $q<1.68$ or $q>5.1$, varying $h$ has no effect. In the intermediate range, increasing $h$ induces transitions such as III $\rightarrow$ IV, II $\rightarrow$ III, or I $\rightarrow$ VIII. Increasing $\beta_1$ to $1$ [Fig.~\ref{fig11}(b)] expands Regime II and introduces Regime V, while Regimes I, III, IV, and VIII shrink.

\begin{figure}
    \centering
    \includegraphics[width=\linewidth]{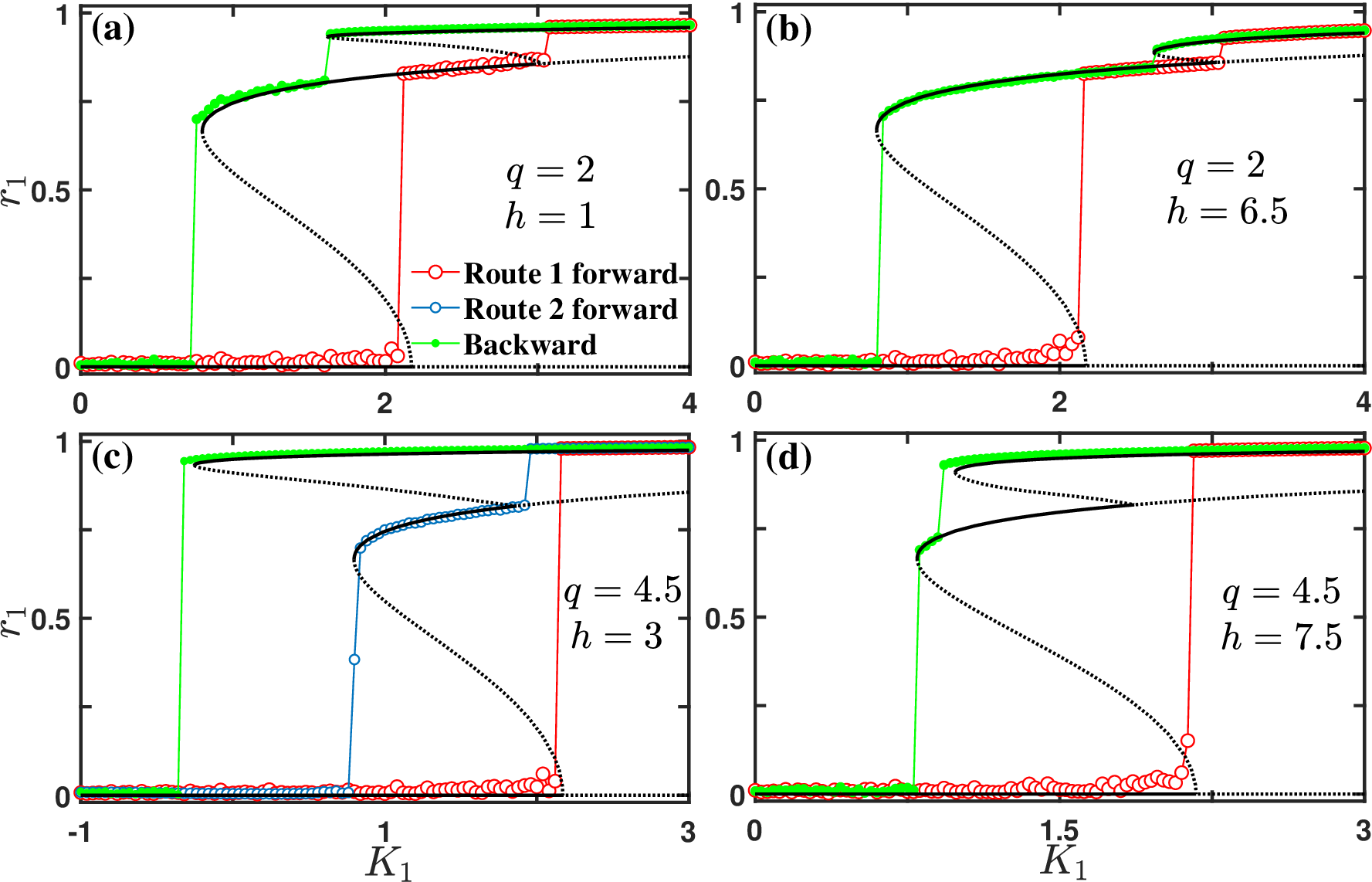}
    \caption{Effect of $h$ on the synchronization transition of $r_1$ at two fixed values of $q$, for $K_{2}=7$, $p=2$, $\beta_1=0.4$, and $\beta_2=1.3$. Bifurcation diagrams of $r_1$ vs. $K_1$ are shown for $q=2$ [(a) $h=1$, (b) $h=6.5$] and $q=4.5$ [(c) $h=3$, (d) $h=7.5$]. At $q=2$, increasing $h$ transforms the forward and backward double jump transitions from a single hysteresis loop at $h=1$ (a) into two distinct hysteresis loops at $h=6.5$ (b). At the larger $q=4.5$, the increase in $h$  changes the multiple, initial-condition-dependent transitions at $h=3$ (c) into an the asymmetric scenario of a single forward jump against a double backward jump at $h=7.5$ (d), corresponding to the newly identified region VIII of Fig.~\ref{fig11}. Black solid and dotted curves denote stable and unstable branches of the reduced-order model; red open and green filled circles denote forward and backward adiabatic continuations, respectively; light-blue open circles in (c) indicate the additional forward route under multiple transitions.}
    \label{fig12}
\end{figure}

To further elucidate the role of $h$, we select four representative points from Fig.~\ref{fig11}(a): $(q,h)=(2,1)$, $(2,6.5)$, $(4.5,3)$, and $(4.5,7.5)$. The corresponding bifurcation diagrams are shown in Figs.~\ref{fig12}(a)--\ref{fig12}(d). For $q=2$ and $h=1$ [Fig.~\ref{fig12}(a)], the system exhibits forward and backward double jump transitions with a single hysteresis loop, whereas increasing $h$ to $6.5$ [Fig.~\ref{fig12}(b)] separates the two explosive transitions, resulting in two distinct hysteresis loops in both directions. For $q=4.5$ and $h=3$ [Fig.~\ref{fig12}(c)], the system exhibits multiple transitions dependent on initial conditions. Increasing $h$ to $7.5$ [Fig.~\ref{fig12}(d)] yields a single forward jump and two backward jumps. These observations are fully consistent with the phase diagram in Fig.~\ref{fig11}(a), and the analytical results show excellent agreement with the numerical simulations.

\begin{figure}
    \centering
    \includegraphics[width=\linewidth]{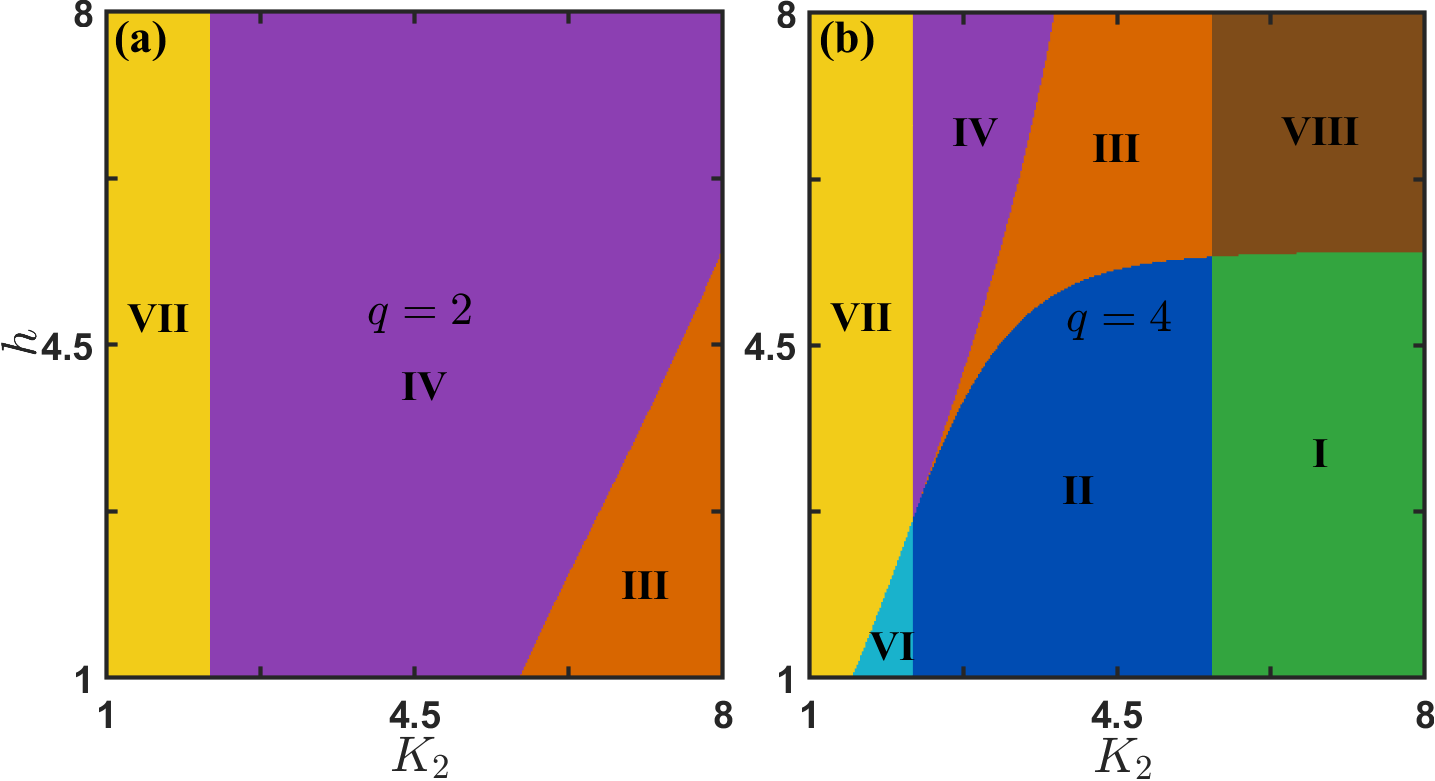}
    \caption{Combined effects of $K_2$, $h$, and $q$ on the transition regimes of $r_1$, obtained from $K_1$ sweeps in the reduced-order model, for $p=2$, $\beta_1=0.4$, and $\beta_2=1.3$. Phase diagrams in the $K_2\text{-}h$ plane are shown for (a) $q=2$ and (b) $q=4$. Roman numerals and colors follow Fig.~\ref{fig2}(a) for regions I--VII and Fig.~\ref{fig11} for region VIII. For $q=2$, only regions III, IV, and VII are present; at $q=4$, regions I, II, VI, and VIII emerge, while IV and VII shrink and III widens.}
    \label{fig13}
\end{figure}

We now turn to the combined effects of $K_2$ and $h$ on the transition scenarios. Figure~\ref{fig13} presents surface plots of the transition regimes in the $K_2\text{-}h$ plane for two values of $q=2$ and $4$, with fixed $p=2$, $\beta_1=0.4$, and $\beta_2=1.3$. For $q=2$ [Fig.~\ref{fig13}(a)], three regimes appear: III, IV, and VII. For fixed $h$, increasing $K_2$ drives the system through sequences VII $\rightarrow$ IV $\rightarrow$ III or VII $\rightarrow$ IV, depending on $h$. For $K_2<5.7$, varying $h$ has no effect. Beyond this threshold, increasing $h$ transforms Regime III to IV. Increasing $q$ to $4$ [Fig.~\ref{fig13}(b)] introduces Regimes I, II, VI, and VIII, while Regimes IV and VII shrink and Regime III expands slightly. In this case, for $K_2<1.5$, varying $h$ has no effect. For larger $K_2$, increasing $h$ yields sequences such as VI $\rightarrow$ VII, II $\rightarrow$ III $\rightarrow$ IV, II $\rightarrow$ IV, or I $\rightarrow$ VIII, depending on the value of $K_2$.

\begin{figure}
    \centering
    \includegraphics[width=\linewidth]{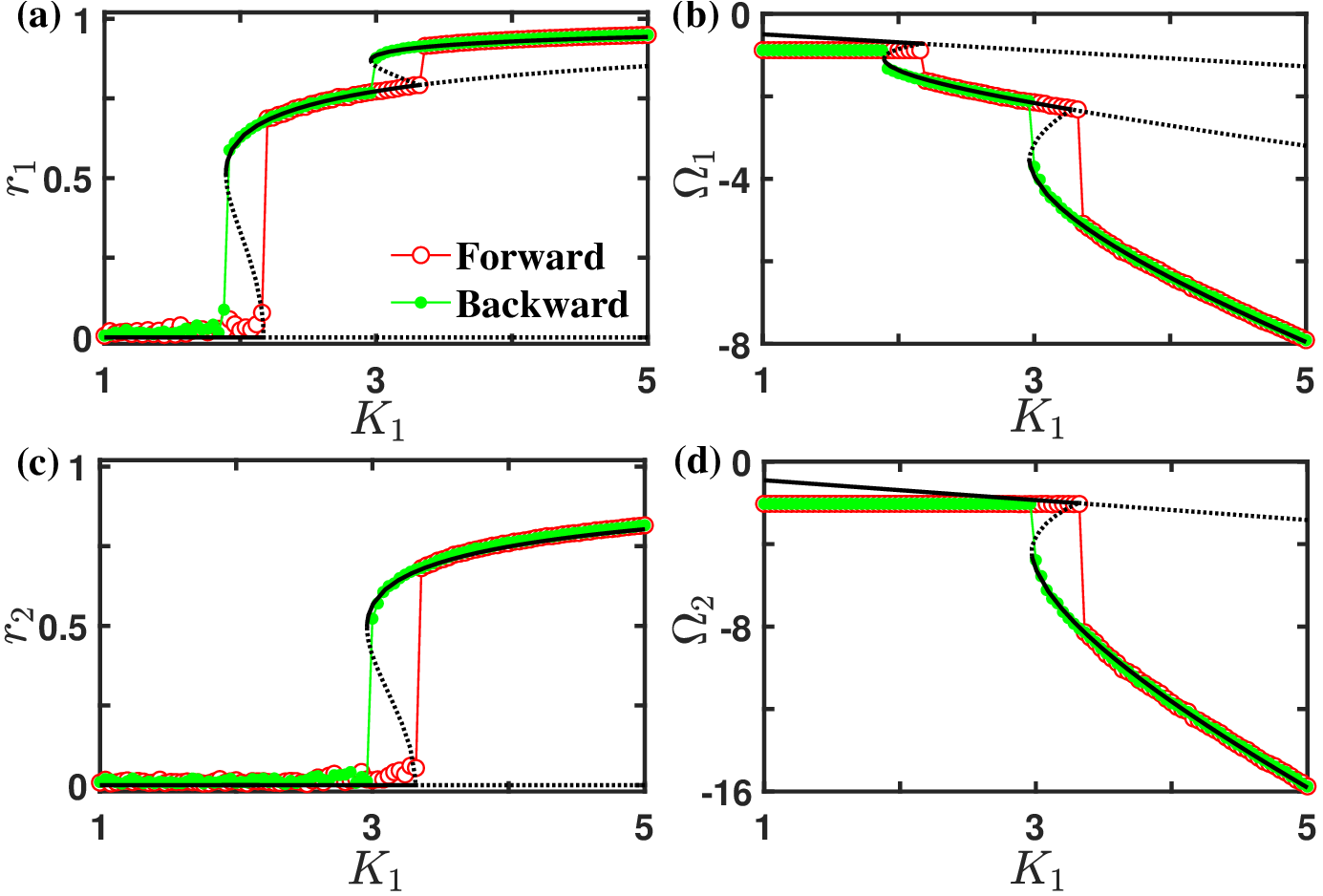}
    \caption{Order parameters and mean frequencies of both layers vs. $K_1$, for $K_{2}=4$, $p=2$, $h=2$, $q=2$, $\beta_1=0.4$, and $\beta_2=1.3$. (a),(b) $r_1$ and $\Omega_1$ for layer 1, showing a double explosive transition; (c),(d) $r_2$ and $\Omega_2$ for layer 2, showing a classical explosive transition. Black solid and dotted curves denote the stable and unstable branches of the reduced-order model, respectively; red open circles and green filled circles denote the forward and backward adiabatic continuations, respectively.}
    \label{fig14}
\end{figure}

Having systematically explored the transition scenarios across various parameter planes, a natural question arises: do the mean-field frequencies of the oscillators reflect the same synchronization transitions observed in the order parameters? Figure~\ref{fig14} addresses this by presenting the variations of the order parameters $r_l$ and the corresponding mean-field frequencies $\Omega_l$ as functions of $K_1$ for fixed $K_{2}=4$, $p=2$, $h=2$, $q=2$, $\beta_1=0.4$, and $\beta_2=1.3$. Panels (a) and (b) show $r_1$ and $\Omega_1$, respectively, while panels (c) and (d) show $r_2$ and $\Omega_2$. In layer 1, both $r_1$ and $\Omega_1$ undergo a double explosive transition, consistent with the behavior observed in the bifurcation diagrams. In layer 2, both $r_2$ and $\Omega_2$ exhibit a classical explosive transition to synchronization. The numerical results for all four quantities are in excellent agreement with the analytical predictions from the reduced-order model.


\section{Conclusions}
This work establishes that the asymmetry between two interacting networks is not merely a perturbation but a generative mechanism for complex synchronization phenomena. We demonstrate double explosive transitions in the forward direction, the backward direction, or a combination of both, accompanied by either single or double hysteresis loops in an adaptive bilayer multiplex network with higher-order interactions and asymmetric phase lags. This behavior arises from the interplay between phase-lag asymmetry $(\beta_1\neq\beta_2)$, cross-layer adaptation, and the higher-order interaction strength $K_2$, which together generate two distinct coherent branches---weak and strong---separated by bifurcations. By systematically charting the transition scenarios across multiple parameter planes, we identify eight distinct synchronization regimes. The occurrence of each regime is governed by the relative ordering of saddle-node and pitchfork bifurcation points, which we derive analytically from the stability analysis.

A central result of our work is that phase-lag asymmetry serves as a robust control knob for tuning the collective dynamics. When the phase lags are symmetric, explosive transitions are suppressed, and hysteresis narrows~\cite{das2026effect}. In contrast, introducing layer-specific phase-lag differences actively promotes multistability and double explosive transitions for fixed higher-order interaction strength $K_2$. Varying $\beta_1$ with fixed $\beta_2$ drives layer 1 from a single explosive jump to double explosive transitions in such regimes, while layer 2 exhibits the mirror behavior under the interchange $\beta_1\leftrightarrow\beta_2$. The phase diagram for $r_2$ is the mirror image of that for $r_1$ across the line $\beta_1=\beta_2$, confirming the layer symmetry of the model. The higher-order interaction strength $K_2$ and the adaptation strengths $q$, $p$, and $h$ further modulate the transition scenarios, revealing that the system is highly tunable. Different parameter combinations can switch the system between distinct synchronization regimes, highlighting that no single parameter acts independently; rather, the combined values of all parameters determine the observed dynamics. The cross-layer nature of the adaptation---where each layer's coupling is modulated by the coherence of the opposite layer---remains essential to the rich synchronization scenarios observed across all regimes. Beyond the specific architecture considered, the principles uncovered here may be relevant to neural systems, where asymmetric coupling between brain regions could modulate pathological synchronization, or to social systems, where subgroup differences in responsiveness can trigger sudden collective shifts. Future work could extend this approach to multilayer networks with more than two layers and higher-order interactions beyond triadic, potentially revealing cascades of multiple explosive transitions across hierarchical levels.



\begin{thebibliography}{55}%
	\makeatletter
	\providecommand \@ifxundefined [1]{%
		\@ifx{#1\undefined}
	}%
	\providecommand \@ifnum [1]{%
		\ifnum #1\expandafter \@firstoftwo
		\else \expandafter \@secondoftwo
		\fi
	}%
	\providecommand \@ifx [1]{%
		\ifx #1\expandafter \@firstoftwo
		\else \expandafter \@secondoftwo
		\fi
	}%
	\providecommand \natexlab [1]{#1}%
	\providecommand \enquote  [1]{``#1''}%
	\providecommand \bibnamefont  [1]{#1}%
	\providecommand \bibfnamefont [1]{#1}%
	\providecommand \citenamefont [1]{#1}%
	\providecommand \href@noop [0]{\@secondoftwo}%
	\providecommand \href [0]{\begingroup \@sanitize@url \@href}%
	\providecommand \@href[1]{\@@startlink{#1}\@@href}%
	\providecommand \@@href[1]{\endgroup#1\@@endlink}%
	\providecommand \@sanitize@url [0]{\catcode `\\12\catcode `\$12\catcode
		`\&12\catcode `\#12\catcode `\^12\catcode `\_12\catcode `\%12\relax}%
	\providecommand \@@startlink[1]{}%
	\providecommand \@@endlink[0]{}%
	\providecommand \url  [0]{\begingroup\@sanitize@url \@url }%
	\providecommand \@url [1]{\endgroup\@href {#1}{\urlprefix }}%
	\providecommand \urlprefix  [0]{URL }%
	\providecommand \Eprint [0]{\href }%
	\providecommand \doibase [0]{https://doi.org/}%
	\providecommand \selectlanguage [0]{\@gobble}%
	\providecommand \bibinfo  [0]{\@secondoftwo}%
	\providecommand \bibfield  [0]{\@secondoftwo}%
	\providecommand \translation [1]{[#1]}%
	\providecommand \BibitemOpen [0]{}%
	\providecommand \bibitemStop [0]{}%
	\providecommand \bibitemNoStop [0]{.\EOS\space}%
	\providecommand \EOS [0]{\spacefactor3000\relax}%
	\providecommand \BibitemShut  [1]{\csname bibitem#1\endcsname}%
	\let\auto@bib@innerbib\@empty
	\bibitem [{\citenamefont {Strogatz}(2001)}]{strogatz2001exploring}%
	\BibitemOpen
	\bibfield  {author} {\bibinfo {author} {\bibfnamefont {S.~H.}\ \bibnamefont
			{Strogatz}},\ }\href@noop {} {\bibfield  {journal} {\bibinfo  {journal}
			{Nature}\ }\textbf {\bibinfo {volume} {410}},\ \bibinfo {pages} {268}
		(\bibinfo {year} {2001})}\BibitemShut {NoStop}%
	\bibitem [{\citenamefont {Newman}(2003)}]{newman2003structure}%
	\BibitemOpen
	\bibfield  {author} {\bibinfo {author} {\bibfnamefont {M.~E.}\ \bibnamefont
			{Newman}},\ }\href@noop {} {\bibfield  {journal} {\bibinfo  {journal} {SIAM
				Review}\ }\textbf {\bibinfo {volume} {45}},\ \bibinfo {pages} {167} (\bibinfo
		{year} {2003})}\BibitemShut {NoStop}%
	\bibitem [{\citenamefont {Boccaletti}\ \emph {et~al.}(2006)\citenamefont
		{Boccaletti}, \citenamefont {Latora}, \citenamefont {Moreno}, \citenamefont
		{Chavez},\ and\ \citenamefont {Hwang}}]{boccaletti2006complex}%
	\BibitemOpen
	\bibfield  {author} {\bibinfo {author} {\bibfnamefont {S.}~\bibnamefont
			{Boccaletti}}, \bibinfo {author} {\bibfnamefont {V.}~\bibnamefont {Latora}},
		\bibinfo {author} {\bibfnamefont {Y.}~\bibnamefont {Moreno}}, \bibinfo
		{author} {\bibfnamefont {M.}~\bibnamefont {Chavez}},\ and\ \bibinfo {author}
		{\bibfnamefont {D.-U.}\ \bibnamefont {Hwang}},\ }\href@noop {} {\bibfield
		{journal} {\bibinfo  {journal} {Physics Reports}\ }\textbf {\bibinfo {volume}
			{424}},\ \bibinfo {pages} {175} (\bibinfo {year} {2006})}\BibitemShut
	{NoStop}%
	\bibitem [{\citenamefont {Christensen}\ and\ \citenamefont
		{Moloney}(2005)}]{christensen2005complexity}%
	\BibitemOpen
	\bibfield  {author} {\bibinfo {author} {\bibfnamefont {K.}~\bibnamefont
			{Christensen}}\ and\ \bibinfo {author} {\bibfnamefont {N.~R.}\ \bibnamefont
			{Moloney}},\ }\href@noop {} {\emph {\bibinfo {title} {Complexity and
				Criticality}}},\ Vol.~\bibinfo {volume} {1}\ (\bibinfo  {publisher} {World
		Scientific Publishing Company},\ \bibinfo {year} {2005})\BibitemShut
	{NoStop}%
	\bibitem [{\citenamefont {Pikovsky}\ \emph {et~al.}(2003)\citenamefont
		{Pikovsky}, \citenamefont {Rosenblum},\ and\ \citenamefont
		{Kurths}}]{pikovsky2003universal}%
	\BibitemOpen
	\bibfield  {author} {\bibinfo {author} {\bibfnamefont {A.}~\bibnamefont
			{Pikovsky}}, \bibinfo {author} {\bibfnamefont {M.}~\bibnamefont
			{Rosenblum}},\ and\ \bibinfo {author} {\bibfnamefont {J.}~\bibnamefont
			{Kurths}},\ }\href@noop {} {\emph {\bibinfo {title} {Synchronization: A
				Universal Concept in Nonlinear Sciences}}},\ Vol.~\bibinfo {volume} {12}\
	(\bibinfo  {publisher} {Cambridge University Press, Cambridge},\ \bibinfo
	{year} {2003})\BibitemShut {NoStop}%
	\bibitem [{\citenamefont {Arenas}\ \emph {et~al.}(2008)\citenamefont {Arenas},
		\citenamefont {D{\'\i}az-Guilera}, \citenamefont {Kurths}, \citenamefont
		{Moreno},\ and\ \citenamefont {Zhou}}]{arenas2008synchronization}%
	\BibitemOpen
	\bibfield  {author} {\bibinfo {author} {\bibfnamefont {A.}~\bibnamefont
			{Arenas}}, \bibinfo {author} {\bibfnamefont {A.}~\bibnamefont
			{D{\'\i}az-Guilera}}, \bibinfo {author} {\bibfnamefont {J.}~\bibnamefont
			{Kurths}}, \bibinfo {author} {\bibfnamefont {Y.}~\bibnamefont {Moreno}},\
		and\ \bibinfo {author} {\bibfnamefont {C.}~\bibnamefont {Zhou}},\ }\href@noop
	{} {\bibfield  {journal} {\bibinfo  {journal} {Physics Reports}\ }\textbf
		{\bibinfo {volume} {469}},\ \bibinfo {pages} {93} (\bibinfo {year}
		{2008})}\BibitemShut {NoStop}%
	\bibitem [{\citenamefont {Pastor-Satorras}\ \emph {et~al.}(2015)\citenamefont
		{Pastor-Satorras}, \citenamefont {Castellano}, \citenamefont {Van~Mieghem},\
		and\ \citenamefont {Vespignani}}]{pastor2015epidemic}%
	\BibitemOpen
	\bibfield  {author} {\bibinfo {author} {\bibfnamefont {R.}~\bibnamefont
			{Pastor-Satorras}}, \bibinfo {author} {\bibfnamefont {C.}~\bibnamefont
			{Castellano}}, \bibinfo {author} {\bibfnamefont {P.}~\bibnamefont
			{Van~Mieghem}},\ and\ \bibinfo {author} {\bibfnamefont {A.}~\bibnamefont
			{Vespignani}},\ }\href@noop {} {\bibfield  {journal} {\bibinfo  {journal}
			{Reviews of Modern Physics}\ }\textbf {\bibinfo {volume} {87}},\ \bibinfo
		{pages} {925} (\bibinfo {year} {2015})}\BibitemShut {NoStop}%
	\bibitem [{\citenamefont {Szab{\'o}}\ and\ \citenamefont
		{Fath}(2007)}]{szabo2007evolutionary}%
	\BibitemOpen
	\bibfield  {author} {\bibinfo {author} {\bibfnamefont {G.}~\bibnamefont
			{Szab{\'o}}}\ and\ \bibinfo {author} {\bibfnamefont {G.}~\bibnamefont
			{Fath}},\ }\href@noop {} {\bibfield  {journal} {\bibinfo  {journal} {Physics
				Reports}\ }\textbf {\bibinfo {volume} {446}},\ \bibinfo {pages} {97}
		(\bibinfo {year} {2007})}\BibitemShut {NoStop}%
	\bibitem [{\citenamefont {Boccaletti}\ \emph {et~al.}(2016)\citenamefont
		{Boccaletti}, \citenamefont {Almendral}, \citenamefont {Guan}, \citenamefont
		{Leyva}, \citenamefont {Liu}, \citenamefont {Sendi{\~n}a-Nadal},
		\citenamefont {Wang},\ and\ \citenamefont {Zou}}]{boccaletti2016explosive}%
	\BibitemOpen
	\bibfield  {author} {\bibinfo {author} {\bibfnamefont {S.}~\bibnamefont
			{Boccaletti}}, \bibinfo {author} {\bibfnamefont {J.~A.}\ \bibnamefont
			{Almendral}}, \bibinfo {author} {\bibfnamefont {S.}~\bibnamefont {Guan}},
		\bibinfo {author} {\bibfnamefont {I.}~\bibnamefont {Leyva}}, \bibinfo
		{author} {\bibfnamefont {Z.}~\bibnamefont {Liu}}, \bibinfo {author}
		{\bibfnamefont {I.}~\bibnamefont {Sendi{\~n}a-Nadal}}, \bibinfo {author}
		{\bibfnamefont {Z.}~\bibnamefont {Wang}},\ and\ \bibinfo {author}
		{\bibfnamefont {Y.}~\bibnamefont {Zou}},\ }\href@noop {} {\bibfield
		{journal} {\bibinfo  {journal} {Physics Reports}\ }\textbf {\bibinfo {volume}
			{660}},\ \bibinfo {pages} {1} (\bibinfo {year} {2016})}\BibitemShut {NoStop}%
	\bibitem [{\citenamefont {D'Souza}\ \emph {et~al.}(2019)\citenamefont
		{D'Souza}, \citenamefont {G{\'o}mez-Gardenes}, \citenamefont {Nagler},\ and\
		\citenamefont {Arenas}}]{d2019explosive}%
	\BibitemOpen
	\bibfield  {author} {\bibinfo {author} {\bibfnamefont {R.~M.}\ \bibnamefont
			{D'Souza}}, \bibinfo {author} {\bibfnamefont {J.}~\bibnamefont
			{G{\'o}mez-Gardenes}}, \bibinfo {author} {\bibfnamefont {J.}~\bibnamefont
			{Nagler}},\ and\ \bibinfo {author} {\bibfnamefont {A.}~\bibnamefont
			{Arenas}},\ }\href@noop {} {\bibfield  {journal} {\bibinfo  {journal}
			{Advances in Physics}\ }\textbf {\bibinfo {volume} {68}},\ \bibinfo {pages}
		{123} (\bibinfo {year} {2019})}\BibitemShut {NoStop}%
	\bibitem [{\citenamefont {G{\'o}mez-Gardenes}\ \emph
		{et~al.}(2011)\citenamefont {G{\'o}mez-Gardenes}, \citenamefont {G{\'o}mez},
		\citenamefont {Arenas},\ and\ \citenamefont {Moreno}}]{gomez2011explosive}%
	\BibitemOpen
	\bibfield  {author} {\bibinfo {author} {\bibfnamefont {J.}~\bibnamefont
			{G{\'o}mez-Gardenes}}, \bibinfo {author} {\bibfnamefont {S.}~\bibnamefont
			{G{\'o}mez}}, \bibinfo {author} {\bibfnamefont {A.}~\bibnamefont {Arenas}},\
		and\ \bibinfo {author} {\bibfnamefont {Y.}~\bibnamefont {Moreno}},\
	}\href@noop {} {\bibfield  {journal} {\bibinfo  {journal} {Physical Review
				Letters}\ }\textbf {\bibinfo {volume} {106}},\ \bibinfo {pages} {128701}
		(\bibinfo {year} {2011})}\BibitemShut {NoStop}%
	\bibitem [{\citenamefont {Leyva}\ \emph {et~al.}(2012)\citenamefont {Leyva},
		\citenamefont {Sevilla-Escoboza}, \citenamefont {Buld{\'u}}, \citenamefont
		{Sendina-Nadal}, \citenamefont {G{\'o}mez-Gardenes}, \citenamefont {Arenas},
		\citenamefont {Moreno}, \citenamefont {Gomez}, \citenamefont
		{Jaimes-Reategui},\ and\ \citenamefont {Boccaletti}}]{leyva2012explosive}%
	\BibitemOpen
	\bibfield  {author} {\bibinfo {author} {\bibfnamefont {I.}~\bibnamefont
			{Leyva}}, \bibinfo {author} {\bibfnamefont {R.}~\bibnamefont
			{Sevilla-Escoboza}}, \bibinfo {author} {\bibfnamefont {J.}~\bibnamefont
			{Buld{\'u}}}, \bibinfo {author} {\bibfnamefont {I.}~\bibnamefont
			{Sendina-Nadal}}, \bibinfo {author} {\bibfnamefont {J.}~\bibnamefont
			{G{\'o}mez-Gardenes}}, \bibinfo {author} {\bibfnamefont {A.}~\bibnamefont
			{Arenas}}, \bibinfo {author} {\bibfnamefont {Y.}~\bibnamefont {Moreno}},
		\bibinfo {author} {\bibfnamefont {S.}~\bibnamefont {Gomez}}, \bibinfo
		{author} {\bibfnamefont {R.}~\bibnamefont {Jaimes-Reategui}},\ and\ \bibinfo
		{author} {\bibfnamefont {S.}~\bibnamefont {Boccaletti}},\ }\href@noop {}
	{\bibfield  {journal} {\bibinfo  {journal} {Physical Review Letters}\
		}\textbf {\bibinfo {volume} {108}},\ \bibinfo {pages} {168702} (\bibinfo
		{year} {2012})}\BibitemShut {NoStop}%
	\bibitem [{\citenamefont {Arola-Fern{\'a}ndez}\ \emph
		{et~al.}(2022)\citenamefont {Arola-Fern{\'a}ndez}, \citenamefont
		{Faci-L{\'a}zaro}, \citenamefont {Skardal}, \citenamefont {Boghiu},
		\citenamefont {G{\'o}mez-Garde{\~n}es},\ and\ \citenamefont
		{Arenas}}]{arola2022emergence}%
	\BibitemOpen
	\bibfield  {author} {\bibinfo {author} {\bibfnamefont {L.}~\bibnamefont
			{Arola-Fern{\'a}ndez}}, \bibinfo {author} {\bibfnamefont {S.}~\bibnamefont
			{Faci-L{\'a}zaro}}, \bibinfo {author} {\bibfnamefont {P.~S.}\ \bibnamefont
			{Skardal}}, \bibinfo {author} {\bibfnamefont {E.-C.}\ \bibnamefont {Boghiu}},
		\bibinfo {author} {\bibfnamefont {J.}~\bibnamefont
			{G{\'o}mez-Garde{\~n}es}},\ and\ \bibinfo {author} {\bibfnamefont
			{A.}~\bibnamefont {Arenas}},\ }\href@noop {} {\bibfield  {journal} {\bibinfo
			{journal} {Communications Physics}\ }\textbf {\bibinfo {volume} {5}},\
		\bibinfo {pages} {264} (\bibinfo {year} {2022})}\BibitemShut {NoStop}%
	\bibitem [{\citenamefont {Kundu}\ \emph {et~al.}(2017)\citenamefont {Kundu},
		\citenamefont {Khanra}, \citenamefont {Hens},\ and\ \citenamefont
		{Pal}}]{kundu2017transition}%
	\BibitemOpen
	\bibfield  {author} {\bibinfo {author} {\bibfnamefont {P.}~\bibnamefont
			{Kundu}}, \bibinfo {author} {\bibfnamefont {P.}~\bibnamefont {Khanra}},
		\bibinfo {author} {\bibfnamefont {C.}~\bibnamefont {Hens}},\ and\ \bibinfo
		{author} {\bibfnamefont {P.}~\bibnamefont {Pal}},\ }\href@noop {} {\bibfield
		{journal} {\bibinfo  {journal} {Physical Review E}\ }\textbf {\bibinfo
			{volume} {96}},\ \bibinfo {pages} {052216} (\bibinfo {year}
		{2017})}\BibitemShut {NoStop}%
	\bibitem [{\citenamefont {Pinto}\ and\ \citenamefont
		{Saa}(2015)}]{pinto2015explosive}%
	\BibitemOpen
	\bibfield  {author} {\bibinfo {author} {\bibfnamefont {R.~S.}\ \bibnamefont
			{Pinto}}\ and\ \bibinfo {author} {\bibfnamefont {A.}~\bibnamefont {Saa}},\
	}\href@noop {} {\bibfield  {journal} {\bibinfo  {journal} {Physical Review
				E}\ }\textbf {\bibinfo {volume} {91}},\ \bibinfo {pages} {022818} (\bibinfo
		{year} {2015})}\BibitemShut {NoStop}%
	\bibitem [{\citenamefont {Kundu}\ and\ \citenamefont
		{Pal}(2019)}]{kundu2019synchronization}%
	\BibitemOpen
	\bibfield  {author} {\bibinfo {author} {\bibfnamefont {P.}~\bibnamefont
			{Kundu}}\ and\ \bibinfo {author} {\bibfnamefont {P.}~\bibnamefont {Pal}},\
	}\href@noop {} {\bibfield  {journal} {\bibinfo  {journal} {Chaos: An
				Interdisciplinary Journal of Nonlinear Science}\ }\textbf {\bibinfo {volume}
			{29}},\ \bibinfo {pages} {013123} (\bibinfo {year} {2019})}\BibitemShut
	{NoStop}%
	\bibitem [{\citenamefont {Zhang}\ \emph {et~al.}(2013)\citenamefont {Zhang},
		\citenamefont {Hu}, \citenamefont {Kurths},\ and\ \citenamefont
		{Liu}}]{zhang2013explosive}%
	\BibitemOpen
	\bibfield  {author} {\bibinfo {author} {\bibfnamefont {X.}~\bibnamefont
			{Zhang}}, \bibinfo {author} {\bibfnamefont {X.}~\bibnamefont {Hu}}, \bibinfo
		{author} {\bibfnamefont {J.}~\bibnamefont {Kurths}},\ and\ \bibinfo {author}
		{\bibfnamefont {Z.}~\bibnamefont {Liu}},\ }\href@noop {} {\bibfield
		{journal} {\bibinfo  {journal} {Physical Review E}\ }\textbf {\bibinfo
			{volume} {88}},\ \bibinfo {pages} {010802} (\bibinfo {year}
		{2013})}\BibitemShut {NoStop}%
	\bibitem [{\citenamefont {Leyva}\ \emph {et~al.}(2013)\citenamefont {Leyva},
		\citenamefont {Sendina-Nadal}, \citenamefont {Almendral}, \citenamefont
		{Navas}, \citenamefont {Olmi},\ and\ \citenamefont
		{Boccaletti}}]{leyva2013explosive}%
	\BibitemOpen
	\bibfield  {author} {\bibinfo {author} {\bibfnamefont {I.}~\bibnamefont
			{Leyva}}, \bibinfo {author} {\bibfnamefont {I.}~\bibnamefont
			{Sendina-Nadal}}, \bibinfo {author} {\bibfnamefont {J.}~\bibnamefont
			{Almendral}}, \bibinfo {author} {\bibfnamefont {A.}~\bibnamefont {Navas}},
		\bibinfo {author} {\bibfnamefont {S.}~\bibnamefont {Olmi}},\ and\ \bibinfo
		{author} {\bibfnamefont {S.}~\bibnamefont {Boccaletti}},\ }\href@noop {}
	{\bibfield  {journal} {\bibinfo  {journal} {Physical Review E}\ }\textbf
		{\bibinfo {volume} {88}},\ \bibinfo {pages} {042808} (\bibinfo {year}
		{2013})}\BibitemShut {NoStop}%
	\bibitem [{\citenamefont {Xu}\ \emph {et~al.}(2016)\citenamefont {Xu},
		\citenamefont {Sun}, \citenamefont {Gao}, \citenamefont {Qiu}, \citenamefont
		{Zheng},\ and\ \citenamefont {Guan}}]{xu2016synchronization}%
	\BibitemOpen
	\bibfield  {author} {\bibinfo {author} {\bibfnamefont {C.}~\bibnamefont
			{Xu}}, \bibinfo {author} {\bibfnamefont {Y.}~\bibnamefont {Sun}}, \bibinfo
		{author} {\bibfnamefont {J.}~\bibnamefont {Gao}}, \bibinfo {author}
		{\bibfnamefont {T.}~\bibnamefont {Qiu}}, \bibinfo {author} {\bibfnamefont
			{Z.}~\bibnamefont {Zheng}},\ and\ \bibinfo {author} {\bibfnamefont
			{S.}~\bibnamefont {Guan}},\ }\href@noop {} {\bibfield  {journal} {\bibinfo
			{journal} {Scientific Reports}\ }\textbf {\bibinfo {volume} {6}},\ \bibinfo
		{pages} {21926} (\bibinfo {year} {2016})}\BibitemShut {NoStop}%
	\bibitem [{\citenamefont {Zhang}\ \emph {et~al.}(2015)\citenamefont {Zhang},
		\citenamefont {Boccaletti}, \citenamefont {Guan},\ and\ \citenamefont
		{Liu}}]{zhang2015explosive}%
	\BibitemOpen
	\bibfield  {author} {\bibinfo {author} {\bibfnamefont {X.}~\bibnamefont
			{Zhang}}, \bibinfo {author} {\bibfnamefont {S.}~\bibnamefont {Boccaletti}},
		\bibinfo {author} {\bibfnamefont {S.}~\bibnamefont {Guan}},\ and\ \bibinfo
		{author} {\bibfnamefont {Z.}~\bibnamefont {Liu}},\ }\href@noop {} {\bibfield
		{journal} {\bibinfo  {journal} {Physical Review Letters}\ }\textbf {\bibinfo
			{volume} {114}},\ \bibinfo {pages} {038701} (\bibinfo {year}
		{2015})}\BibitemShut {NoStop}%
	\bibitem [{\citenamefont {Danziger}\ \emph {et~al.}(2016)\citenamefont
		{Danziger}, \citenamefont {Moskalenko}, \citenamefont {Kurkin}, \citenamefont
		{Zhang}, \citenamefont {Havlin},\ and\ \citenamefont
		{Boccaletti}}]{danziger2016explosive}%
	\BibitemOpen
	\bibfield  {author} {\bibinfo {author} {\bibfnamefont {M.~M.}\ \bibnamefont
			{Danziger}}, \bibinfo {author} {\bibfnamefont {O.~I.}\ \bibnamefont
			{Moskalenko}}, \bibinfo {author} {\bibfnamefont {S.~A.}\ \bibnamefont
			{Kurkin}}, \bibinfo {author} {\bibfnamefont {X.}~\bibnamefont {Zhang}},
		\bibinfo {author} {\bibfnamefont {S.}~\bibnamefont {Havlin}},\ and\ \bibinfo
		{author} {\bibfnamefont {S.}~\bibnamefont {Boccaletti}},\ }\href@noop {}
	{\bibfield  {journal} {\bibinfo  {journal} {Chaos: An Interdisciplinary
				Journal of Nonlinear Science}\ }\textbf {\bibinfo {volume} {26}},\ \bibinfo
		{pages} {065307} (\bibinfo {year} {2016})}\BibitemShut {NoStop}%
	\bibitem [{\citenamefont {Khanra}\ \emph {et~al.}(2018)\citenamefont {Khanra},
		\citenamefont {Kundu}, \citenamefont {Hens},\ and\ \citenamefont
		{Pal}}]{khanra2018explosive}%
	\BibitemOpen
	\bibfield  {author} {\bibinfo {author} {\bibfnamefont {P.}~\bibnamefont
			{Khanra}}, \bibinfo {author} {\bibfnamefont {P.}~\bibnamefont {Kundu}},
		\bibinfo {author} {\bibfnamefont {C.}~\bibnamefont {Hens}},\ and\ \bibinfo
		{author} {\bibfnamefont {P.}~\bibnamefont {Pal}},\ }\href@noop {} {\bibfield
		{journal} {\bibinfo  {journal} {Physical Review E}\ }\textbf {\bibinfo
			{volume} {98}},\ \bibinfo {pages} {052315} (\bibinfo {year}
		{2018})}\BibitemShut {NoStop}%
	\bibitem [{\citenamefont {Khanra}\ and\ \citenamefont
		{Pal}(2021)}]{khanra2021explosive}%
	\BibitemOpen
	\bibfield  {author} {\bibinfo {author} {\bibfnamefont {P.}~\bibnamefont
			{Khanra}}\ and\ \bibinfo {author} {\bibfnamefont {P.}~\bibnamefont {Pal}},\
	}\href@noop {} {\bibfield  {journal} {\bibinfo  {journal} {Chaos, Solitons \&
				Fractals}\ }\textbf {\bibinfo {volume} {143}},\ \bibinfo {pages} {110621}
		(\bibinfo {year} {2021})}\BibitemShut {NoStop}%
	\bibitem [{\citenamefont {Peron}\ and\ \citenamefont
		{Rodrigues}(2012)}]{peron2012explosive}%
	\BibitemOpen
	\bibfield  {author} {\bibinfo {author} {\bibfnamefont {T.~K.~D.}\
			\bibnamefont {Peron}}\ and\ \bibinfo {author} {\bibfnamefont {F.~A.}\
			\bibnamefont {Rodrigues}},\ }\href@noop {} {\bibfield  {journal} {\bibinfo
			{journal} {Physical Review E}\ }\textbf {\bibinfo {volume} {86}},\ \bibinfo
		{pages} {016102} (\bibinfo {year} {2012})}\BibitemShut {NoStop}%
	\bibitem [{\citenamefont {Kachhvah}\ and\ \citenamefont
		{Jalan}(2019)}]{kachhvah2019delay}%
	\BibitemOpen
	\bibfield  {author} {\bibinfo {author} {\bibfnamefont {A.~D.}\ \bibnamefont
			{Kachhvah}}\ and\ \bibinfo {author} {\bibfnamefont {S.}~\bibnamefont
			{Jalan}},\ }\href@noop {} {\bibfield  {journal} {\bibinfo  {journal} {New
				Journal of Physics}\ }\textbf {\bibinfo {volume} {21}},\ \bibinfo {pages}
		{015006} (\bibinfo {year} {2019})}\BibitemShut {NoStop}%
	\bibitem [{\citenamefont {Skardal}\ and\ \citenamefont
		{Arenas}(2020)}]{skardal2020higher}%
	\BibitemOpen
	\bibfield  {author} {\bibinfo {author} {\bibfnamefont {P.~S.}\ \bibnamefont
			{Skardal}}\ and\ \bibinfo {author} {\bibfnamefont {A.}~\bibnamefont
			{Arenas}},\ }\href@noop {} {\bibfield  {journal} {\bibinfo  {journal}
			{Communications Physics}\ }\textbf {\bibinfo {volume} {3}},\ \bibinfo {pages}
		{218} (\bibinfo {year} {2020})}\BibitemShut {NoStop}%
	\bibitem [{\citenamefont {Laptyeva}\ \emph {et~al.}(2025)\citenamefont
		{Laptyeva}, \citenamefont {Jalan},\ and\ \citenamefont
		{Ivanchenko}}]{laptyeva2025explosive}%
	\BibitemOpen
	\bibfield  {author} {\bibinfo {author} {\bibfnamefont {T.}~\bibnamefont
			{Laptyeva}}, \bibinfo {author} {\bibfnamefont {S.}~\bibnamefont {Jalan}},\
		and\ \bibinfo {author} {\bibfnamefont {M.}~\bibnamefont {Ivanchenko}},\
	}\href@noop {} {\bibfield  {journal} {\bibinfo  {journal} {Chaos, Solitons \&
				Fractals}\ }\textbf {\bibinfo {volume} {192}},\ \bibinfo {pages} {116003}
		(\bibinfo {year} {2025})}\BibitemShut {NoStop}%
	\bibitem [{\citenamefont {Malizia}\ \emph {et~al.}(2025)\citenamefont
		{Malizia}, \citenamefont {Lamata-Ot{\'\i}n}, \citenamefont {Frasca},
		\citenamefont {Latora},\ and\ \citenamefont
		{G{\'o}mez-Garde{\~n}es}}]{malizia2025hyperedge}%
	\BibitemOpen
	\bibfield  {author} {\bibinfo {author} {\bibfnamefont {F.}~\bibnamefont
			{Malizia}}, \bibinfo {author} {\bibfnamefont {S.}~\bibnamefont
			{Lamata-Ot{\'\i}n}}, \bibinfo {author} {\bibfnamefont {M.}~\bibnamefont
			{Frasca}}, \bibinfo {author} {\bibfnamefont {V.}~\bibnamefont {Latora}},\
		and\ \bibinfo {author} {\bibfnamefont {J.}~\bibnamefont
			{G{\'o}mez-Garde{\~n}es}},\ }\href@noop {} {\bibfield  {journal} {\bibinfo
			{journal} {Nature Communications}\ }\textbf {\bibinfo {volume} {16}},\
		\bibinfo {pages} {555} (\bibinfo {year} {2025})}\BibitemShut {NoStop}%
	\bibitem [{\citenamefont {Sakaguchi}\ and\ \citenamefont
		{Kuramoto}(1986)}]{sakaguchi1986soluble}%
	\BibitemOpen
	\bibfield  {author} {\bibinfo {author} {\bibfnamefont {H.}~\bibnamefont
			{Sakaguchi}}\ and\ \bibinfo {author} {\bibfnamefont {Y.}~\bibnamefont
			{Kuramoto}},\ }\href@noop {} {\bibfield  {journal} {\bibinfo  {journal}
			{Progress of Theoretical Physics}\ }\textbf {\bibinfo {volume} {76}},\
		\bibinfo {pages} {576} (\bibinfo {year} {1986})}\BibitemShut {NoStop}%
	\bibitem [{\citenamefont {Omel’Chenko}\ and\ \citenamefont
		{Wolfrum}(2012)}]{omel2012nonuniversal}%
	\BibitemOpen
	\bibfield  {author} {\bibinfo {author} {\bibfnamefont {O.~E.}\ \bibnamefont
			{Omel’Chenko}}\ and\ \bibinfo {author} {\bibfnamefont {M.}~\bibnamefont
			{Wolfrum}},\ }\href@noop {} {\bibfield  {journal} {\bibinfo  {journal}
			{Physical Review Letters}\ }\textbf {\bibinfo {volume} {109}},\ \bibinfo
		{pages} {164101} (\bibinfo {year} {2012})}\BibitemShut {NoStop}%
	\bibitem [{\citenamefont {Omel’chenko}\ and\ \citenamefont
		{Wolfrum}(2013)}]{omel2013bifurcations}%
	\BibitemOpen
	\bibfield  {author} {\bibinfo {author} {\bibfnamefont {E.}~\bibnamefont
			{Omel’chenko}}\ and\ \bibinfo {author} {\bibfnamefont {M.}~\bibnamefont
			{Wolfrum}},\ }\href@noop {} {\bibfield  {journal} {\bibinfo  {journal}
			{Physica D: Nonlinear Phenomena}\ }\textbf {\bibinfo {volume} {263}},\
		\bibinfo {pages} {74} (\bibinfo {year} {2013})}\BibitemShut {NoStop}%
	\bibitem [{\citenamefont {Omel'chenko}\ and\ \citenamefont
		{Wolfrum}(2016)}]{omel2016there}%
	\BibitemOpen
	\bibfield  {author} {\bibinfo {author} {\bibfnamefont {O.~E.}\ \bibnamefont
			{Omel'chenko}}\ and\ \bibinfo {author} {\bibfnamefont {M.}~\bibnamefont
			{Wolfrum}},\ }\href@noop {} {\bibfield  {journal} {\bibinfo  {journal}
			{Chaos: An Interdisciplinary Journal of Nonlinear Science}\ }\textbf
		{\bibinfo {volume} {26}},\ \bibinfo {pages} {094806} (\bibinfo {year}
		{2016})}\BibitemShut {NoStop}%
	\bibitem [{\citenamefont {Skardal}\ and\ \citenamefont
		{Xu}(2022)}]{skardal2022tiered}%
	\BibitemOpen
	\bibfield  {author} {\bibinfo {author} {\bibfnamefont {P.~S.}\ \bibnamefont
			{Skardal}}\ and\ \bibinfo {author} {\bibfnamefont {C.}~\bibnamefont {Xu}},\
	}\href@noop {} {\bibfield  {journal} {\bibinfo  {journal} {Chaos: An
				Interdisciplinary Journal of Nonlinear Science}\ }\textbf {\bibinfo {volume}
			{32}},\ \bibinfo {pages} {053120} (\bibinfo {year} {2022})}\BibitemShut
	{NoStop}%
	\bibitem [{\citenamefont {Rajwani}\ \emph {et~al.}(2023)\citenamefont
		{Rajwani}, \citenamefont {Suman},\ and\ \citenamefont
		{Jalan}}]{rajwani2023tiered}%
	\BibitemOpen
	\bibfield  {author} {\bibinfo {author} {\bibfnamefont {P.}~\bibnamefont
			{Rajwani}}, \bibinfo {author} {\bibfnamefont {A.}~\bibnamefont {Suman}},\
		and\ \bibinfo {author} {\bibfnamefont {S.}~\bibnamefont {Jalan}},\
	}\href@noop {} {\bibfield  {journal} {\bibinfo  {journal} {Chaos: An
				Interdisciplinary Journal of Nonlinear Science}\ }\textbf {\bibinfo {volume}
			{33}},\ \bibinfo {pages} {061102} (\bibinfo {year} {2023})}\BibitemShut
	{NoStop}%
	\bibitem [{\citenamefont {Manoranjani}\ \emph {et~al.}(2023)\citenamefont
		{Manoranjani}, \citenamefont {Saiprasad}, \citenamefont {Gopal},
		\citenamefont {Senthilkumar},\ and\ \citenamefont
		{Chandrasekar}}]{manoranjani2023phase}%
	\BibitemOpen
	\bibfield  {author} {\bibinfo {author} {\bibfnamefont {M.}~\bibnamefont
			{Manoranjani}}, \bibinfo {author} {\bibfnamefont {V.}~\bibnamefont
			{Saiprasad}}, \bibinfo {author} {\bibfnamefont {R.}~\bibnamefont {Gopal}},
		\bibinfo {author} {\bibfnamefont {D.}~\bibnamefont {Senthilkumar}},\ and\
		\bibinfo {author} {\bibfnamefont {V.}~\bibnamefont {Chandrasekar}},\
	}\href@noop {} {\bibfield  {journal} {\bibinfo  {journal} {Physical Review
				E}\ }\textbf {\bibinfo {volume} {108}},\ \bibinfo {pages} {044307} (\bibinfo
		{year} {2023})}\BibitemShut {NoStop}%
	\bibitem [{\citenamefont {Das}\ \emph {et~al.}(2026)\citenamefont {Das},
		\citenamefont {Dutta},\ and\ \citenamefont {Pal}}]{das2026effect}%
	\BibitemOpen
	\bibfield  {author} {\bibinfo {author} {\bibfnamefont {A.~B.}\ \bibnamefont
			{Das}}, \bibinfo {author} {\bibfnamefont {S.}~\bibnamefont {Dutta}},\ and\
		\bibinfo {author} {\bibfnamefont {P.}~\bibnamefont {Pal}},\ }\href@noop {}
	{\bibfield  {journal} {\bibinfo  {journal} {Physical Review E}\ }\textbf
		{\bibinfo {volume} {113}},\ \bibinfo {pages} {044303} (\bibinfo {year}
		{2026})}\BibitemShut {NoStop}%
	\bibitem [{\citenamefont {Olmi}\ \emph {et~al.}(2014)\citenamefont {Olmi},
		\citenamefont {Navas}, \citenamefont {Boccaletti},\ and\ \citenamefont
		{Torcini}}]{olmi2014hysteretic}%
	\BibitemOpen
	\bibfield  {author} {\bibinfo {author} {\bibfnamefont {S.}~\bibnamefont
			{Olmi}}, \bibinfo {author} {\bibfnamefont {A.}~\bibnamefont {Navas}},
		\bibinfo {author} {\bibfnamefont {S.}~\bibnamefont {Boccaletti}},\ and\
		\bibinfo {author} {\bibfnamefont {A.}~\bibnamefont {Torcini}},\ }\href@noop
	{} {\bibfield  {journal} {\bibinfo  {journal} {Physical Review E}\ }\textbf
		{\bibinfo {volume} {90}},\ \bibinfo {pages} {042905} (\bibinfo {year}
		{2014})}\BibitemShut {NoStop}%
	\bibitem [{\citenamefont {Gao}\ and\ \citenamefont
		{Efstathiou}(2021)}]{gao2021synchronized}%
	\BibitemOpen
	\bibfield  {author} {\bibinfo {author} {\bibfnamefont {J.}~\bibnamefont
			{Gao}}\ and\ \bibinfo {author} {\bibfnamefont {K.}~\bibnamefont
			{Efstathiou}},\ }\href@noop {} {\bibfield  {journal} {\bibinfo  {journal}
			{Chaos: An Interdisciplinary Journal of Nonlinear Science}\ }\textbf
		{\bibinfo {volume} {31}},\ \bibinfo {pages} {093137} (\bibinfo {year}
		{2021})}\BibitemShut {NoStop}%
	\bibitem [{\citenamefont {Carballosa}\ \emph {et~al.}(2023)\citenamefont
		{Carballosa}, \citenamefont {Mu{\~n}uzuri}, \citenamefont {Boccaletti},
		\citenamefont {Torcini},\ and\ \citenamefont {Olmi}}]{carballosa2023cluster}%
	\BibitemOpen
	\bibfield  {author} {\bibinfo {author} {\bibfnamefont {A.}~\bibnamefont
			{Carballosa}}, \bibinfo {author} {\bibfnamefont {A.~P.}\ \bibnamefont
			{Mu{\~n}uzuri}}, \bibinfo {author} {\bibfnamefont {S.}~\bibnamefont
			{Boccaletti}}, \bibinfo {author} {\bibfnamefont {A.}~\bibnamefont
			{Torcini}},\ and\ \bibinfo {author} {\bibfnamefont {S.}~\bibnamefont
			{Olmi}},\ }\href@noop {} {\bibfield  {journal} {\bibinfo  {journal} {Chaos,
				Solitons \& Fractals}\ }\textbf {\bibinfo {volume} {177}},\ \bibinfo {pages}
		{114197} (\bibinfo {year} {2023})}\BibitemShut {NoStop}%
	\bibitem [{\citenamefont {Fialkowski}\ \emph {et~al.}(2023)\citenamefont
		{Fialkowski}, \citenamefont {Yanchuk}, \citenamefont {Sokolov}, \citenamefont
		{Sch{\"o}ll}, \citenamefont {Gottwald},\ and\ \citenamefont
		{Berner}}]{fialkowski2023heterogeneous}%
	\BibitemOpen
	\bibfield  {author} {\bibinfo {author} {\bibfnamefont {J.}~\bibnamefont
			{Fialkowski}}, \bibinfo {author} {\bibfnamefont {S.}~\bibnamefont {Yanchuk}},
		\bibinfo {author} {\bibfnamefont {I.~M.}\ \bibnamefont {Sokolov}}, \bibinfo
		{author} {\bibfnamefont {E.}~\bibnamefont {Sch{\"o}ll}}, \bibinfo {author}
		{\bibfnamefont {G.~A.}\ \bibnamefont {Gottwald}},\ and\ \bibinfo {author}
		{\bibfnamefont {R.}~\bibnamefont {Berner}},\ }\href@noop {} {\bibfield
		{journal} {\bibinfo  {journal} {Physical Review Letters}\ }\textbf {\bibinfo
			{volume} {130}},\ \bibinfo {pages} {067402} (\bibinfo {year}
		{2023})}\BibitemShut {NoStop}%
	\bibitem [{\citenamefont {Seif}\ and\ \citenamefont
		{Zarei}(2025)}]{seif2025double}%
	\BibitemOpen
	\bibfield  {author} {\bibinfo {author} {\bibfnamefont {A.}~\bibnamefont
			{Seif}}\ and\ \bibinfo {author} {\bibfnamefont {M.}~\bibnamefont {Zarei}},\
	}\href@noop {} {\bibfield  {journal} {\bibinfo  {journal} {Chaos, Solitons \&
				Fractals}\ }\textbf {\bibinfo {volume} {196}},\ \bibinfo {pages} {116412}
		(\bibinfo {year} {2025})}\BibitemShut {NoStop}%
	\bibitem [{\citenamefont {Li}\ \emph {et~al.}(2025)\citenamefont {Li},
		\citenamefont {Pal}, \citenamefont {Lei}, \citenamefont {Ghosh},\ and\
		\citenamefont {Small}}]{li2025higher}%
	\BibitemOpen
	\bibfield  {author} {\bibinfo {author} {\bibfnamefont {X.}~\bibnamefont
			{Li}}, \bibinfo {author} {\bibfnamefont {P.~K.}\ \bibnamefont {Pal}},
		\bibinfo {author} {\bibfnamefont {Y.}~\bibnamefont {Lei}}, \bibinfo {author}
		{\bibfnamefont {D.}~\bibnamefont {Ghosh}},\ and\ \bibinfo {author}
		{\bibfnamefont {M.}~\bibnamefont {Small}},\ }\href@noop {} {\bibfield
		{journal} {\bibinfo  {journal} {Physical Review E}\ }\textbf {\bibinfo
			{volume} {111}},\ \bibinfo {pages} {024303} (\bibinfo {year}
		{2025})}\BibitemShut {NoStop}%
	\bibitem [{\citenamefont {Dutta}\ \emph {et~al.}(2025)\citenamefont {Dutta},
		\citenamefont {Kundu}, \citenamefont {Khanra}, \citenamefont {Minati},
		\citenamefont {Boccaletti}, \citenamefont {Pal},\ and\ \citenamefont
		{Hens}}]{dutta2025double}%
	\BibitemOpen
	\bibfield  {author} {\bibinfo {author} {\bibfnamefont {S.}~\bibnamefont
			{Dutta}}, \bibinfo {author} {\bibfnamefont {P.}~\bibnamefont {Kundu}},
		\bibinfo {author} {\bibfnamefont {P.}~\bibnamefont {Khanra}}, \bibinfo
		{author} {\bibfnamefont {L.}~\bibnamefont {Minati}}, \bibinfo {author}
		{\bibfnamefont {S.}~\bibnamefont {Boccaletti}}, \bibinfo {author}
		{\bibfnamefont {P.}~\bibnamefont {Pal}},\ and\ \bibinfo {author}
		{\bibfnamefont {C.}~\bibnamefont {Hens}},\ }\href@noop {} {\bibfield
		{journal} {\bibinfo  {journal} {Physical Review Research}\ }\textbf {\bibinfo
			{volume} {7}},\ \bibinfo {pages} {L022049} (\bibinfo {year}
		{2025})}\BibitemShut {NoStop}%
	\bibitem [{\citenamefont {Costa}\ \emph {et~al.}(2025)\citenamefont {Costa},
		\citenamefont {Novaes},\ and\ \citenamefont {de~Aguiar}}]{costa2025exact}%
	\BibitemOpen
	\bibfield  {author} {\bibinfo {author} {\bibfnamefont {G.~S.}\ \bibnamefont
			{Costa}}, \bibinfo {author} {\bibfnamefont {M.}~\bibnamefont {Novaes}},\ and\
		\bibinfo {author} {\bibfnamefont {M.~A.}\ \bibnamefont {de~Aguiar}},\
	}\href@noop {} {\bibfield  {journal} {\bibinfo  {journal} {Chaos, Solitons \&
				Fractals}\ }\textbf {\bibinfo {volume} {195}},\ \bibinfo {pages} {116243}
		(\bibinfo {year} {2025})}\BibitemShut {NoStop}%
	\bibitem [{\citenamefont {Buld{\'u}}\ and\ \citenamefont
		{Porter}(2018)}]{buldu2018frequency}%
	\BibitemOpen
	\bibfield  {author} {\bibinfo {author} {\bibfnamefont {J.~M.}\ \bibnamefont
			{Buld{\'u}}}\ and\ \bibinfo {author} {\bibfnamefont {M.~A.}\ \bibnamefont
			{Porter}},\ }\href@noop {} {\bibfield  {journal} {\bibinfo  {journal}
			{Network Neuroscience}\ }\textbf {\bibinfo {volume} {2}},\ \bibinfo {pages}
		{418} (\bibinfo {year} {2018})}\BibitemShut {NoStop}%
	\bibitem [{\citenamefont {Buzs{\'a}ki}(2006)}]{buzsaki2006rhythms}%
	\BibitemOpen
	\bibfield  {author} {\bibinfo {author} {\bibfnamefont {G.}~\bibnamefont
			{Buzs{\'a}ki}},\ }\href@noop {} {\emph {\bibinfo {title} {Rhythms of the
				Brain}}}\ (\bibinfo  {publisher} {Oxford University Press, Oxford},\ \bibinfo
	{year} {2006})\BibitemShut {NoStop}%
	\bibitem [{\citenamefont {B{\"o}ttcher}\ \emph {et~al.}(2020)\citenamefont
		{B{\"o}ttcher}, \citenamefont {Otto}, \citenamefont {Kettemann},\ and\
		\citenamefont {Agert}}]{bottcher2020time}%
	\BibitemOpen
	\bibfield  {author} {\bibinfo {author} {\bibfnamefont {P.~C.}\ \bibnamefont
			{B{\"o}ttcher}}, \bibinfo {author} {\bibfnamefont {A.}~\bibnamefont {Otto}},
		\bibinfo {author} {\bibfnamefont {S.}~\bibnamefont {Kettemann}},\ and\
		\bibinfo {author} {\bibfnamefont {C.}~\bibnamefont {Agert}},\ }\href@noop {}
	{\bibfield  {journal} {\bibinfo  {journal} {Chaos: An Interdisciplinary
				Journal of Nonlinear Science}\ }\textbf {\bibinfo {volume} {30}},\ \bibinfo
		{pages} {013122} (\bibinfo {year} {2020})}\BibitemShut {NoStop}%
	\bibitem [{\citenamefont {Wang}\ \emph
		{et~al.}(2017{\natexlab{a}})\citenamefont {Wang}, \citenamefont {Pumir},
		\citenamefont {Garnier},\ and\ \citenamefont {Liu}}]{wang2017explosive}%
	\BibitemOpen
	\bibfield  {author} {\bibinfo {author} {\bibfnamefont {C.-Q.}\ \bibnamefont
			{Wang}}, \bibinfo {author} {\bibfnamefont {A.}~\bibnamefont {Pumir}},
		\bibinfo {author} {\bibfnamefont {N.~B.}\ \bibnamefont {Garnier}},\ and\
		\bibinfo {author} {\bibfnamefont {Z.-H.}\ \bibnamefont {Liu}},\ }\href@noop
	{} {\bibfield  {journal} {\bibinfo  {journal} {Frontiers of Physics}\
		}\textbf {\bibinfo {volume} {12}},\ \bibinfo {pages} {128901} (\bibinfo
		{year} {2017}{\natexlab{a}})}\BibitemShut {NoStop}%
	\bibitem [{\citenamefont {Wang}\ \emph
		{et~al.}(2017{\natexlab{b}})\citenamefont {Wang}, \citenamefont {Tian},
		\citenamefont {Dhamala},\ and\ \citenamefont {Liu}}]{wang2017small}%
	\BibitemOpen
	\bibfield  {author} {\bibinfo {author} {\bibfnamefont {Z.}~\bibnamefont
			{Wang}}, \bibinfo {author} {\bibfnamefont {C.}~\bibnamefont {Tian}}, \bibinfo
		{author} {\bibfnamefont {M.}~\bibnamefont {Dhamala}},\ and\ \bibinfo {author}
		{\bibfnamefont {Z.}~\bibnamefont {Liu}},\ }\href@noop {} {\bibfield
		{journal} {\bibinfo  {journal} {Scientific Reports}\ }\textbf {\bibinfo
			{volume} {7}},\ \bibinfo {pages} {561} (\bibinfo {year}
		{2017}{\natexlab{b}})}\BibitemShut {NoStop}%
	\bibitem [{\citenamefont {Dobson}\ \emph {et~al.}(2007)\citenamefont {Dobson},
		\citenamefont {Carreras}, \citenamefont {Lynch},\ and\ \citenamefont
		{Newman}}]{dobson2007complex}%
	\BibitemOpen
	\bibfield  {author} {\bibinfo {author} {\bibfnamefont {I.}~\bibnamefont
			{Dobson}}, \bibinfo {author} {\bibfnamefont {B.~A.}\ \bibnamefont
			{Carreras}}, \bibinfo {author} {\bibfnamefont {V.~E.}\ \bibnamefont
			{Lynch}},\ and\ \bibinfo {author} {\bibfnamefont {D.~E.}\ \bibnamefont
			{Newman}},\ }\href@noop {} {\bibfield  {journal} {\bibinfo  {journal} {Chaos:
				An Interdisciplinary Journal of Nonlinear Science}\ }\textbf {\bibinfo
			{volume} {17}},\ \bibinfo {pages} {026103} (\bibinfo {year}
		{2007})}\BibitemShut {NoStop}%
	\bibitem [{\citenamefont {Filatrella}\ \emph {et~al.}(2007)\citenamefont
		{Filatrella}, \citenamefont {Pedersen},\ and\ \citenamefont
		{Wiesenfeld}}]{filatrella2007generalized}%
	\BibitemOpen
	\bibfield  {author} {\bibinfo {author} {\bibfnamefont {G.}~\bibnamefont
			{Filatrella}}, \bibinfo {author} {\bibfnamefont {N.~F.}\ \bibnamefont
			{Pedersen}},\ and\ \bibinfo {author} {\bibfnamefont {K.}~\bibnamefont
			{Wiesenfeld}},\ }\href@noop {} {\bibfield  {journal} {\bibinfo  {journal}
			{Physical Review E}\ }\textbf {\bibinfo {volume} {75}},\ \bibinfo {pages}
		{017201} (\bibinfo {year} {2007})}\BibitemShut {NoStop}%
	\bibitem [{\citenamefont {Zou}\ and\ \citenamefont
		{Wang}(2020)}]{zou2020dynamics}%
	\BibitemOpen
	\bibfield  {author} {\bibinfo {author} {\bibfnamefont {W.}~\bibnamefont
			{Zou}}\ and\ \bibinfo {author} {\bibfnamefont {J.}~\bibnamefont {Wang}},\
	}\href@noop {} {\bibfield  {journal} {\bibinfo  {journal} {Physical Review
				E}\ }\textbf {\bibinfo {volume} {102}},\ \bibinfo {pages} {012219} (\bibinfo
		{year} {2020})}\BibitemShut {NoStop}%
	\bibitem [{\citenamefont {Xu}\ \emph {et~al.}(2021)\citenamefont {Xu},
		\citenamefont {Tang}, \citenamefont {L{\"u}}, \citenamefont {Alfaro-Bittner},
		\citenamefont {Boccaletti}, \citenamefont {Perc},\ and\ \citenamefont
		{Guan}}]{xu2021collective}%
	\BibitemOpen
	\bibfield  {author} {\bibinfo {author} {\bibfnamefont {C.}~\bibnamefont
			{Xu}}, \bibinfo {author} {\bibfnamefont {X.}~\bibnamefont {Tang}}, \bibinfo
		{author} {\bibfnamefont {H.}~\bibnamefont {L{\"u}}}, \bibinfo {author}
		{\bibfnamefont {K.}~\bibnamefont {Alfaro-Bittner}}, \bibinfo {author}
		{\bibfnamefont {S.}~\bibnamefont {Boccaletti}}, \bibinfo {author}
		{\bibfnamefont {M.}~\bibnamefont {Perc}},\ and\ \bibinfo {author}
		{\bibfnamefont {S.}~\bibnamefont {Guan}},\ }\href@noop {} {\bibfield
		{journal} {\bibinfo  {journal} {Physical Review Research}\ }\textbf {\bibinfo
			{volume} {3}},\ \bibinfo {pages} {043004} (\bibinfo {year}
		{2021})}\BibitemShut {NoStop}%
	\bibitem [{\citenamefont {Ott}\ and\ \citenamefont
		{Antonsen}(2008)}]{ott2008low}%
	\BibitemOpen
	\bibfield  {author} {\bibinfo {author} {\bibfnamefont {E.}~\bibnamefont
			{Ott}}\ and\ \bibinfo {author} {\bibfnamefont {T.~M.}\ \bibnamefont
			{Antonsen}},\ }\href@noop {} {\bibfield  {journal} {\bibinfo  {journal}
			{Chaos: An Interdisciplinary Journal of Nonlinear Science}\ }\textbf
		{\bibinfo {volume} {18}},\ \bibinfo {pages} {037113} (\bibinfo {year}
		{2008})}\BibitemShut {NoStop}%
	\bibitem [{\citenamefont {Biswas}\ and\ \citenamefont
		{Gupta}(2024)}]{biswas2024effect}%
	\BibitemOpen
	\bibfield  {author} {\bibinfo {author} {\bibfnamefont {D.}~\bibnamefont
			{Biswas}}\ and\ \bibinfo {author} {\bibfnamefont {S.}~\bibnamefont {Gupta}},\
	}\href@noop {} {\bibfield  {journal} {\bibinfo  {journal} {Physical Review
				E}\ }\textbf {\bibinfo {volume} {109}},\ \bibinfo {pages} {024221} (\bibinfo
		{year} {2024})}\BibitemShut {NoStop}%
\end{thebibliography}

%

\end{document}